\documentclass[%
 reprint,
 amsmath,amssymb,
 aps,
prb,
]{revtex4-2}

\usepackage{color}
\usepackage{graphicx}
\usepackage{dcolumn}
\usepackage{bm}
\usepackage{braket}

\begin{document}

\preprint{APS/123-QED}

\title{Nonreciprocal Magnon by Symmetric Anisotropic Exchange Interaction\\
in Honeycomb Antiferromagnet}

\author{Takuya Matsumoto}
\affiliation{%
 Department of Physics, Hokkaido University, Sapporo 060-0810, Japan
 }%
\author{Satoru Hayami}%
 \affiliation{%
 Department of Applied Physics, The University of Tokyo, Tokyo 113-8656, Japan
 }%

\date{\today}

\begin{abstract}
We investigate a microscopic origin of nonreciprocal magnon that is distinct from the Dzyaloshinskii-Moriya interaction in a honeycomb antiferromagnet. 
The key ingredient is a symmetric anisotropic exchange interaction depending on the bond direction, which results in a valley-type nonreciprocal magnon excitations under the staggered antiferromagnetic ordering. 
Furthermore, we find this type of nonreciprocal magnon exhibits a peculiar magnetic-field response; the nonreciprocal direction can be manipulated by the in-plane rotating magnetic field. 
The obtained results can be accounted for the emergence of the magnetic toroidal multipoles.
\end{abstract}

\maketitle


\section{introduction}
\label{intro}

Magnetism in the absence of the spatial inversion symmetry has drawn considerable interest in condensed matter physics, since it exhibits various fascinating phenomena, such as the magneto-electric effect~\cite{Curie, Fiebig_2005, khomskii2009trend}  and nonreciprocal transport~\cite{wakatsuki2017nonreciprocal}. 
For example, magnetic skyrmions in polar/chiral magnets show nonreciprocal directional dichroism due to the lack of the spatial inversion symmetry~\cite{PhysRevB.87.134403}. 
Recently, current-induced magnetization and magneto-piezo electricity in antiferromagnetic (AFM) metals without the inversion symmetry have been observed in experiments~\cite{doi:10.7566/JPSJ.87.033702, magpiezo1, PhysRevLett.122.127207}. 

Such magnets without the spatial inversion symmetry also affect collective excitations of magnon and photon, which results in directional-dependent dynamical properties even in magnetic insulators~\cite{PhysRevLett.30.125, doi:10.1143/JPSJ.56.3635, cortes2013influence, PhysRevB.88.184404, doi:10.7566/JPSJ.85.053705, 10.1038/s41598-019-51646-3, kawano2019designing, PhysRevB.100.174402}.
Theoretically, the nonreciprocal magnons have long been studied in the magnetic systems with the 
Dzyaloshinsky-Moriya (DM) interaction~\cite{DZYALOSHINSKY1958241, PhysRev.120.91}, which were observed in recent experiments~\cite{PhysRevB.92.184419, zhang2015plane, cho2015thickness, PhysRevB.94.144420, PhysRevB.93.235131, PhysRevLett.119.047201, PhysRevB.95.220406,  tacchi2017interfacial, chaurasiya2018dependence, PhysRevB.98.064416}.
Among them, asymmetric (nonreciprocal) magnon dispersions were directly detected in the noncentrosymmetric ferromagnet $\mathrm{LiFe_5O_8}$~\cite{PhysRevB.92.184419} and AFM $\alpha$-$\rm{Cu_2 V_2O_7}$~\cite{PhysRevLett.119.047201} through the spectroscopic measurements.
Furthermore, such asymmetric magnons give rise to directional-dependent physical phenomena,
such as nonreciprocal magneto-optical
~\cite{takahashi.Nat.Phys., doi:10.1143/JPSJ.81.023712, Miyahara2013, PhysRevB.89.195145, PhysRevLett.114.197203, PhysRevB.98.134422, PhysRevB.99.094401} and nonreciprocal spin Seebeck effects~\cite{PhysRevB.98.020401, PhysRevB.96.180414}. 
Meanwhile, some nonreciprocal-magnon mechanisms which are different from the DM interaction have been found, e.g., the dipolar coupling between ferromagnetic layers~\cite{grunberg1986layered, zhang1987spin, di2015enhancement, gallardo2019reconfigurable, albisetti2020optically}, the vector spin chirality in the spiral spin structures~\cite{PhysRevB.89.195145, doi:10.1143/JPSJ.81.023712, takahashi.Nat.Phys., PhysRevB.98.184405}, the bond-dependent symmetric anisotropic exchange interaction~\cite{maksimov2019anisotropic}, and magnetic interactions induced by curved magnetic surfaces~\cite{otalora2016curvature}, and graded magnetization~\cite{gallardo2019spin} in ferromagnetic films.
Among them, the bond-dependent symmetric anisotropic and DM interactions originate from the spin-orbit coupling in bulk systems, which are different from each other: 
The former mechanism does not require the inversion symmetry breaking on the bond center, while the latter does. 
Thus, nonreciprocal magnons can be realized even in centrosymmetric magnets when the magnetic order breaks the inversion symmetry in the presence of the bond-dependent symmetric anisotropic exchange interaction, which will extend the scope of functional materials toward applications to AFM spintronics devices.
Nevertheless, its microscopic mechanism has not been elucidated thus far.

In the present study, we investigate the behavior of nonreciprocal magnons under the anisotropic magnetic interactions on the basis of point group symmetry.
We show that the threefold bond-dependent symmetric exchange interaction in the honeycomb structure leads to a valley-type nonreciprocal magnon excitations once the staggered-type collinear AFM ordering occurs.
We also find that its nonreciprocal magnon excitations show an in-plane angle-dependent directional dispersions under an external magnetic field.
We present that the microscopic origin of the nonreciprocal magnon excitations is attributed to the emergent magnetic toroidal multipoles hidden in the cluster magnetic structure from the symmetry point of view.

The organization of this paper is as follows.
 In Sec.~\ref{model}, we introduce the spin model and outline the linear spin wave calculations based on the Holstein-Primakoff transformations. 
 In Sec.~\ref{result},  we give the nonreciprocal magnon excitations at both zero and nonzero magnetic fields. 
 Section~\ref{summary} is devoted to a summary of the present paper.
 In Appendix~\ref{GS}, we present the calculations of the spin configurations in the ground state.
 In Appendix~\ref{DM}, we compare the magnon dispersions obtained in the present paper with those in the presence of the 
DM interaction. 

\section{model and method}
\label{model}

\begin{figure}[t]
\includegraphics[width=1.0\linewidth]{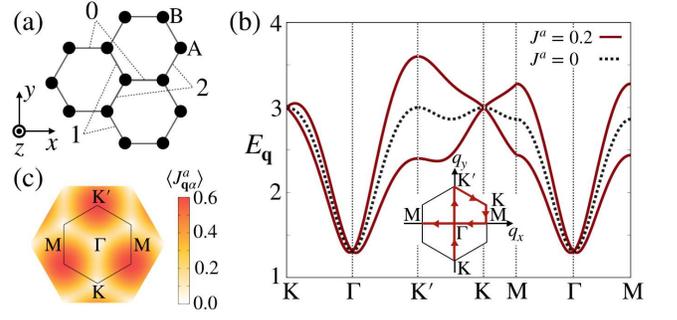}
\caption{\label{fig1} (a) Schematic picture of the honeycomb structure consisting of A and B sublattices. 
The bond index 0-2 is also shown.
(b) The magnon dispersions in the model in Eq.~\eqref{eq4} at $J=0.9$ and $H=0$.  
The red solid (black dotted) lines represent the result at $J^a=0.2$ ($J^a=0$).
In the inset, the first Brillouin zone is shown. 
(c) The color plot of $\langle J^a_{\mathbf{q}\alpha} \rangle$ in Eq.~(\ref{ja}) in $\mathbf{q}$ space.}
\end{figure}

Let us start by considering the localized spin model in the honeycomb structure, as shown in Fig.~\ref{fig1}(a). 
By taking into account the symmetry elements of the honeycomb structure under the point group $6/mmm$, the spin Hamiltonian with the symmetry-allowed exchange interactions is given by 
 \begin{align}
 \label{eq1}
\mathcal{H}=&\sum_{\langle ij \rangle} 
 \Big[
 J(S^+_{i{\rm{A}}}S^-_{j{\rm{B}}}+S^-_{i{\rm{A}}}S^+_{j{\rm{B}}}) +J^z S^z_{i{\rm{A}}} S^z_{j{\rm{B}}} \nonumber \\ 
 &+J^a(\gamma_{ij}S^+_{i{\rm{A}}}S^+_{j{\rm{B}}}+\gamma_{ij}^{*}S^-_{i{\rm{A}}}S^-_{j{\rm{B}}})
 \Big]
 -\sum_{i, \eta}
 \mathbf{H}\cdot \mathbf{S}_{i \eta},
  \end{align}
where $S^{\zeta}_{i \eta}$ is a classical spin with a $\zeta=x, y,z$ component at unit cell $i$ and sublattice $\eta=$ A, B, and $S^{\pm}_{i \eta}\equiv (S^x_{i \eta}\pm iS^y_{i \eta})/\sqrt{2}$.
The sum of $\langle ij \rangle$ is taken for the nearest-neighbor spins. 
The first two terms in 
the square braket in Eq.~\eqref{eq1} represent the $xxz$-type AFM exchange interactions where we assume $J_z >J>0$.
The third term stands for a bond-dependent 
symmetric anisotropic exchange interaction with the coupling constant $J^a$ and the phase factor $\gamma_{ij}\equiv {\rm{e}}^{i\frac{2\pi n}{3}}$ where $n=0,1, 2$ corresponds to the three nearest-neighbor bonds in Fig.~\ref{fig1}(a).
This term originates from the relativistic spin-orbit coupling in multi-orbital systems where the competition between the crystalline electric field and the atomic spin-orbit coupling gives rise to a Kramers doublet under the large total angular momentum, although it is different from the DM interaction which appears in the absence of the inversion symmetry at the bond center.
A similar bond-dependent symmetric anisotropic exchange interaction has recently been studied in the triangle AFM~\cite{PhysRevB.94.035107} and honeycomb ferromagnet~\cite{PhysRevB.95.014435, PhysRevApplied.9.024029}. 
The second term is a Zeeman interaction under an external in-plane magnetic field, $\mathbf{H}=(H_x, H_y, 0)=H(\cos \phi, \sin \phi,0)$.  
We set $J^z=1$ as the energy unit and the distance between A and B sublattices to be 1.

In order to discuss the magnon excitations, we investigate the optimal spin pattern within the two-sublattice orderings in the model in Eq.~(\ref{eq1}). 
For $J^z>J$, the spin configurations are given by $\mathbf{S}_{i{\rm A}}=S(\sin{\theta}\cos{\phi}, \sin{\theta}\sin{\phi}, \cos{\theta})$ and $\mathbf{S}_{i{\rm B}}=S(\sin{\theta}\cos{\phi}, \sin{\theta}\sin{\phi}, -\cos{\theta})$ where $\theta=\sin^{-1}\left[H/3S(J+J^z)\right]$, as shown in Appendix~\ref{GS}. 
Note that $J^a$ does not contribute to the ground-state energy within the two-sublattice AFM ordering, although it plays an important role in asymmetric magnon excitations as discussed in Sec.~\ref{result}.

For the above spin configuration, we examine magnetic excitations by using the linear spin-wave theory. 
We adopt the standard Holstein-Primakoff transformation as $\tilde{S}^{+}_{i\eta}\equiv \sqrt{S}\eta'_i$, $\tilde{S}^{-}_{i\eta}\equiv \sqrt{S}\eta_i'^{\dagger}$, and $\tilde{S}^z_{i\eta}\equiv S-\eta_i'^{\dagger}\eta'_i$ where $S=1$ and $(\tilde{S}^{x}_{i\eta},\tilde{S}^{y}_{i\eta},\tilde{S}^{z}_{i\eta})^T$ is the local rotated frame with the quantization axis along the $\tilde{S}^{z}_{i\eta}$ direction and $\eta'_i=a_i$ and $b_i$ are the boson operator for sublattice $\eta=$ A and B, respectively. 
By performing the Fourier transformation, the spin-wave Hamiltonian in momentum ($\mathbf{q}$) space is obtained as
\begin{align}
\label{eq4}
\mathcal{H}=\frac{1}{2}
\sum_{\mathbf{q}}
\Psi^{\dagger}_{\mathbf{q}}
\begin{pmatrix}
X(\mathbf{q})&Y(\mathbf{q})\\
Y^{*}(-\mathbf{q})&X^{*}(-\mathbf{q})
\end{pmatrix}
\Psi_{\mathbf{q}},
\end{align}
where $\Psi^{\dagger}_{\mathbf{q}}=(a^{\dagger}_{\mathbf{q}},b^{\dagger}_{\mathbf{q}},a_{-\mathbf{q}},b_{-\mathbf{q}})$.
We omit the classical ground-state energy per unit cell $E_{\rm{GS}}=-3J^zS^2-H^2/[3(J^z+J)]$.
 $X(\mathbf{q})$ and $Y(\mathbf{q})$ in Eq.~\eqref{eq4} are $2\times 2$ matrices, which are given by
\begin{align}
\label{eq5}
X(\mathbf{q})=&
\begin{pmatrix}
Z&\sum_{n}F_n{\rm{e}}^{i\mathbf{q} \cdot \boldsymbol{\rho}_n }\\
\sum_{n} F^{*}_n{\rm{e}}^{-i\mathbf{q} \cdot \boldsymbol{\rho}_n}&Z
\end{pmatrix},\\
\label{eq52}
Y(\mathbf{q})=&
\begin{pmatrix}
 0&\sum_{n}G_n{\rm{e}}^{i\mathbf{q} \cdot \boldsymbol{\rho}_n }\\
 \sum_{n}G_n{\rm{e}}^{-i\mathbf{q} \cdot \boldsymbol{\rho}_n }&0
\end{pmatrix},
\end{align}
where the sum of $n$ is taken for the three nearest-neighbor bonds ($n=0,1,2$) with $\boldsymbol{\rho}_0=(1,0)
$, $\boldsymbol{\rho}_1=(-1/2,\sqrt{3}/2)$, and $\boldsymbol{\rho}_2=(-1/2,-\sqrt{3}/2)$.
In Eqs.~\eqref{eq5} and \eqref{eq52}, $F_n$, $G_n$, and $Z$ are expressed as
\begin{align}
F_n=&
\frac{J+J^z}{2}\sin^2{\theta} \nonumber \\
&-J^a\left[\cos{\Phi_n}\frac{1+\cos^2{\theta}}{2} - i  \sin{\Phi_n}\cos{\theta}\right],\\
G_n=&
-J+\sin^2{\theta}\left[ \frac{J+J^z}{2}
+\frac{J^a}{2}\cos{\Phi_n}\right],
\end{align}
and $Z=H\sin\theta-3J\sin^2{\theta}+3J^z\cos^2{\theta}$ where $\Phi_n=2\phi+\chi_n$ and $\chi_n=0, 2\pi/3, 4\pi/3$ for $n=0,1,2$.
We use the numerical Bogoliubov transformation for the Hamiltonian in Eq.~(\ref{eq4}) for the magnon dispersions~\cite{COLPA1978327}.

\section{result} 
\label{result}

In this section, we first show the nonreciprocal magnon excitations under a collinear AFM order in Sec.~\ref{H=0}.
Next, we discuss the nonreciprocal behavior under the external magnetic field in Sec.~\ref{Hneq0}.

\subsection{Collinear antiferromagnetic order at zero field}
\label{H=0}

We show the result in the absence of the magnetic field ($H=0$) where the staggered collinear AFM order with the moments along the $z$ direction, i.e., $\theta=0$, becomes the ground state.
Figure~\ref{fig1}(b) shows the magnon dispersions at $J=0.9$, $J^a=0.2$, and $H=0$ (red solid lines).
For comparison, we also show the magnon dispersions at $J^a=0$ (black dotted lines).
Compared to the result at $J^a=0$, the magnon dispersions at $J^a=0.2$ split in the entire Brillouin zone except for the $\Gamma$ and K points.
The splitting of magnon excitation spectrum is characterized in an asymmetric way: the magnon dispersions undergo an antisymmetric deformation with respect to $\mathbf{q}$ for $J^a \neq 0$.

To examine the effect of $J^a$ on the antisymmetric magnon dispersions, we calculate its contribution by evaluating
the expectation value at each momentum $\mathbf{q}$ in the the third term in the square braket in Eq.~(\ref{eq1}), which is represented by 
\begin{align}
\label{ja}
\langle J^a_{\mathbf{q}\zeta} \rangle \equiv&
J^a \sum_n \bra{\zeta_{\mathbf{q}}}
{\rm{e}}^{i\mathbf{q} \cdot \boldsymbol{\rho}_n }(\bar{F}_n a^{\dagger}_{\mathbf{q}}b_{\mathbf{q}}+\bar{F}_n^* a_{-\mathbf{q}}b^{\dagger}_{-\mathbf{q}}) \nonumber \\
&+{\rm{e}}^{i\mathbf{q} \cdot \boldsymbol{\rho}_n } \bar{G}_n (a^{\dagger}_{\mathbf{q}}b^{\dagger}_{-\mathbf{q}}+a_{-\mathbf{q}}b_{\mathbf{q}}) + {\rm H.c.}
\ket{\zeta_{\mathbf{q}}},
\end{align}
where $\ket{\zeta_{\mathbf{q}}}=\zeta^{\dagger}_{\mathbf{q}}\ket{0}$ stands for the eigenmode where $\zeta_{\mathbf{q}}=\alpha_{\mathbf{q}}$ ($\beta_{\mathbf{q}}$) for the upper (lower) magnon band.
$\bar{F}_n$ and $\bar{G}_n$ are given by $\bar{F}_n=[\cos{\Phi_n}(1+\cos^2{\theta})/2 - i  \sin{\Phi_n}\cos{\theta}]$ and $\bar{G}_n=\cos{\Phi_n}\sin^2{\theta}$.
Figure~\ref{fig1}(c) shows the color plot of $\langle J^a_{\mathbf{q}\alpha}\rangle$ in the entire $\mathbf{q}$ space where $\langle J^a_{\mathbf{q}\alpha}\rangle \simeq -\langle J^a_{\mathbf{q}\beta}\rangle$.
In Fig.~\ref{fig1}(c), $\langle J^a_{\mathbf{q}\alpha}\rangle$ remains a threefold rotational symmetry in the form of $\sin (\sqrt{3} q_y/2) [\cos (3 q_x/2)-\cos (\sqrt{3} q_y/2)]$, which is symmetric along the M-$\Gamma$ line ($q_x \leftrightarrow -q_x$) and asymmetric along the K-$\Gamma$-$\rm{K}'$ line ($q_y \leftrightarrow -q_y$). 
Reflecting such a functional form, $\langle J^a_{\mathbf{q}\alpha}\rangle$ becomes the maximum at the $\rm{K}'$ point.
The behavior of $\langle J^a_{\mathbf{q}\zeta}\rangle$ in Fig.~\ref{fig1}(c) is consistent with the magnon-band splitting in Fig.~\ref{fig1}(b).
In fact, the magnon-band splitting $\Delta E_\mathbf{q}$ is related with $\langle J^a_{\mathbf{q}\zeta}\rangle$ as $\Delta E_\mathbf{q}=\langle J^a_{\mathbf{q}\alpha}\rangle-\langle J^a_{\mathbf{q}\beta}\rangle$.

We analytically evaluate the asymmetric magnon-band splitting $\Delta E_\mathbf{q}$ by using the perturbation analysis with respect to $J^a$. 
The lowest-energy correction by $J^a$ is given by the first-order perturbation, which is obtained as  
\begin{align}
\label{eq6}
\Delta E_\mathbf{q}=|J^a|\sqrt{3+2\left[\sum_{n=0,1,2}\cos\left(\mathbf{q} \cdot \boldsymbol{\rho}'_n+\frac{2\pi}{3}\right)\right]},
\end{align}
where we assume $J^z \gg J$ for simplicity.
In Eq.~(\ref{eq6}), $\boldsymbol{\rho}'_n$ is the next-nearest-neighbor vector as $\boldsymbol{\rho}'_0=\bm{\rho}_0-\bm{\rho}_1$, $\boldsymbol{\rho}'_1=\bm{\rho}_1-\bm{\rho}_2$, and $\boldsymbol{\rho}'_2=\bm{\rho}_2-\bm{\rho}_0$.
Note that Eq.~(\ref{eq6}) describes $\Delta E_\mathbf{q} \neq \Delta E_{-\mathbf{q}}$.
The expression indicates that an effective kinetic motion of magnons between the next-nearest-neighbor spins plays an important role in inducing the antisymmetric magnon-band splitting.
Moreover, Eq.~(\ref{eq6}) shows that the nonreciprocal magnon dispersion is proportional to the symmetric anisotropic exchange $J^a$ and  is irrespective of the sign of $J^a$.

Such an emergent asymmetric magnon structure in $\mathbf{q}$ space is also caused by the DM interaction, which appears without the inversion symmetry at the bond center. 
However, the way of asymmetric spectra is qualitative different from each other, although both of them originate from the spin-orbit coupling microscopically. 
The asymmetric magnon dispersion induced by the symmetric anisotropic exchange interaction exhibits the magnon-band splitting except for the high-symmetry points, $\Gamma$ and K, in Fig.~\ref{fig1}(b), whereas that by the DM interaction does not show any splittings; the twofold degenerated bands at the K and $\rm{K}'$ points move in an opposite direction, as shown in Appendix~\ref{DM}~\cite{doi:10.7566/JPSJ.85.053705}. 
Thus, these two contributions are separably detected by the spectroscopic measurements~\cite{PhysRevB.92.184419,PhysRevLett.119.047201}.  
Moreover, another difference is found in the microscopic origin. 
The key issue for the present mechanism appears in the exchange interactions between the the {\it nearest-neighbor} bonds.
On the other hand, the exchange interactions between the {\it next-nearest-neighbor} bonds give the contribution to nonreciprocal magnons for the mechanisms based on the DM interaction~\cite{doi:10.7566/JPSJ.85.053705}.
Thus, nonreciprocal magnons in the honeycomb AFM can be expected even when the next-nearest-neighbor exchange couplings including the DM interaction are negligibly small.

\begin{figure}[t]
\includegraphics[width=1.0\linewidth]{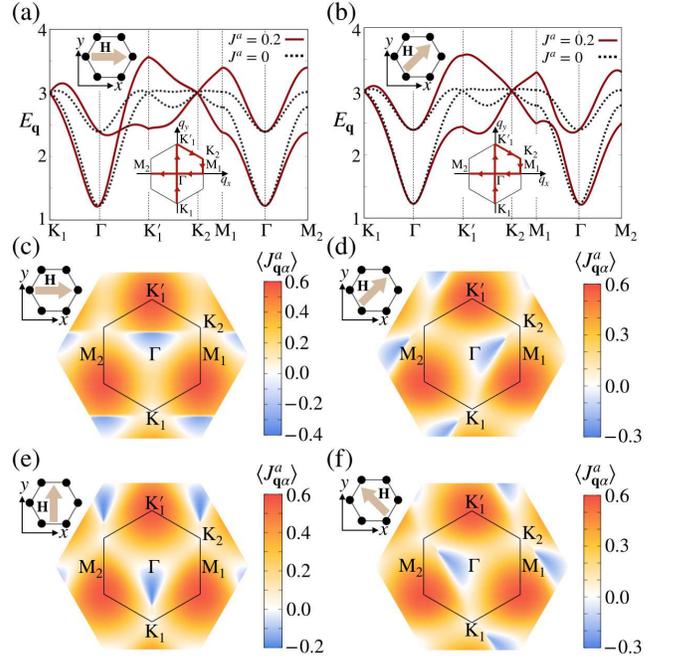}
\caption{
\label{fig2}
(a) and (b) magnon bands at (a) $\mathbf{H}
=(H,0,0)$ ($\phi=0$) and (b) $\mathbf{H}
=(H,H,0)/\sqrt{2}$ ($\phi=\pi/4$) with $H=2$.
The other model parameters in Eq.~\eqref{eq4} are the same as those in Fig.~\ref{fig1}(b). 
(c)-(f) $\langle J^a_{\mathbf{q}\alpha} \rangle$ for (c)$\phi=0$, (d)$\phi=\pi/4$, (e)$\phi=\pi/2$, and (f)$\phi=3\pi/4$. 
}
\end{figure}

\subsection{Canted antiferromagnetic order at nonzero field}
\label{Hneq0}

We discuss an additional asymmetric magnon deformation under the magnetic field ($H \neq 0$).
Figures~\ref{fig2}(a) and \ref{fig2}(b) represent the results in the presence of the in-plane magnetic field $\mathbf{H}=H(\cos \phi, \sin \phi, 0)$ for $\phi=0$ and $\phi=\pi/4$, respectively. 
The magnon dispersions in Figs.~\ref{fig2}(a) and \ref{fig2}(b) are different from each other for $J^a \neq 0$, while they are the same for $J^a=0$. 
For instance, the magnon bands are symmetric (asymmetric) along the M$_1$-$\Gamma$-M$_2$ line in Fig.~\ref{fig2}(a) [Fig.~\ref{fig2}(b)], which means that the antisymmetric functional form depends on the magnetic-field direction.

In order to display the antisymmetric modulations under the in-plane magnetic field, we show $\langle J^a_{\mathbf{q}\alpha} \rangle$ ($\simeq -\langle J^a_{\mathbf{q}\beta} \rangle$) in Eq.~\eqref{ja} for several values of $\phi$: $\phi=0$ in Fig.~\ref{fig2}(c), $\phi=\pi/4$ in Fig.~\ref{fig2}(d), $\phi=\pi/2$ in Fig.~\ref{fig2}(e), and $\phi=3\pi/4$ in Fig.~\ref{fig2}(f).
In contrast to the result at $H=0$ in Fig.~\ref{fig1}(b), $\langle J^a_{\mathbf{q}\alpha} \rangle$ in Figs.~\ref{fig2}(c)-\ref{fig2}(f) breaks the threefold rotational symmetry: there are linearly antisymmetric modulations against $q_y$ along the [100] and [010] field directions ($\phi=0$ and $\phi=\pi/2$) in Figs.~\ref{fig2}(c) and \ref{fig2}(e) and against $q_x$ along the $[110]$ and $[\bar{1}10]$ field directions ($\phi=\pi/4$ and $\phi=3\pi/4$) in Figs.~\ref{fig2}(d) and \ref{fig2}(f).
In contrast to the case at zero field in Sec.~\ref{H=0}, the magnon-band splitting $\Delta E(\mathbf{q})$ is slightly deviated from $\langle J^a_{\mathbf{q}\alpha}\rangle-\langle J^a_{\mathbf{q}\beta}\rangle$ in the presence of $\mathbf{H}$.

\begin{figure}[t]
\includegraphics[width=1.0\linewidth]{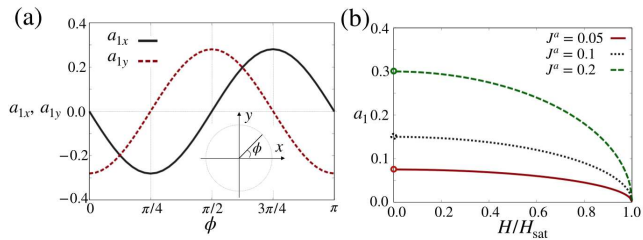}
\caption{\label{fig3}(a) 
Magnetic-field angle dependences of the linear coefficients in the magnon band, $a_{1x}$ and $a_{1y}$, at $J=0.9$, $J^a=0.2$, and $H=2$.  
(b) $H$ dependence of $a_1(\equiv \sqrt{a_{1x}^2+a_{1y}^2})$ for $J^a=0.05$, $0.1$, and $0.2$ where $H_{\rm sat}$ represents the saturated magnetic field.
}
\end{figure}

We analyze additional antisymmetric modulations in the magnon dispersions by considering the $\mathbf{q}\to 0$ limit.
By setting $\mathbf{q}=(q_x, q_y)=q (\cos \phi_q, \sin \phi_q)$, the magnon dispersion for upper band $E_{\mathbf{q} \alpha}$ is expanded as $E_{\mathbf{q} \alpha}=a_0+ (a_{1x} \cos\phi_q+a_{1y} \sin\phi_q)q +\mathcal{O} (\mathbf{q}^2)$ where $a_0$, $a_{1x}$, and $a_{1y}$ are the expansion coefficients.
We show the field-angle dependence of the linear coefficients $a_{1x}$ and $a_{1y}$ obtained by performing the numerical differentiation for upper band in Fig.~\ref{fig3}(a).
As clearly shown in Fig.~\ref{fig3}(a), there are linear antisymmetric 
modulations in the magnon dispersions under the in-plane magnetic field. 
The angle dependences of $a_{1x}$ and $a_{1y}$ are fitted as $-\sin(2\phi)$ and $-\cos(2\phi)$, respectively, where their norm $a_1 \equiv \sqrt{a_{1x}^2+a_{1y}^2}$ is independent of $\phi$.
The linear antisymmetric direction is rotated by $-2\phi$ when the field direction is rotated by $\phi$.

The result suggests that the nonreciprocal dispersion can be controlled by the magnetic-field direction, since the nonreciprocal transport is dominantly characterized by the linear antisymmetric components~\cite{PhysRevB.98.134422, PhysRevB.99.094401, PhysRevB.96.180414}.
Moreover, it is noted that such an angle dependence of nonreciprocal magnon does not occur under the DM interaction that might appear in the next-nearest-neighbor bonds in the honeycomb AFM, as shown in Appendix~\ref{DM}. 
Thus, the symmetric anisotropic exchange-driven nonreciprocal magnon can be detected by measuring the conductive and response tensors in the nonreciprocal magneto-optical~\cite{takahashi.Nat.Phys., doi:10.1143/JPSJ.81.023712, Miyahara2013, PhysRevB.89.195145, PhysRevLett.114.197203, PhysRevB.98.134422, PhysRevB.99.094401} and spin Seebeck effects~\cite{PhysRevB.98.020401, PhysRevB.96.180414} besides the microscopic spectroscopic measurements.

Figure~\ref{fig3}(b)  shows the $H$ dependence of $a_1$ for $J^a=0.05$, $0.1$, and $0.2$ where $H_{\rm sat}$ represents the saturated magnetic field. 
The value of $a_1$ becomes gradually small while increasing the magnetic field, and it vanishes at $H_{\rm sat}$.
Note that there is a finite jump of $a_1$ for infinitesimally small $H$, whose discontinuity is presumably due to the presence of the band crossing at the $\Gamma$ point, as shown in Fig.~\ref{fig1}(b).
The magnitude of the linear coefficient $a_1$ is proportional to $J^a$ in Fig.~\ref{fig3}(b).

Finally, let us discuss the peculiar angle dependence of the nonreciprocal excitations in terms of emergent magnetic toroidal multipoles~\cite{Spaldin_0953-8984-20-43-434203, hayami2018microscopic, PhysRevB.98.165110}. 
In the case of $H=0$ where the staggered AFM state with the moments along the $z$ direction is stabilized, this AFM state is regarded as a ferroic alignment of the odd-parity magnetic toroidal octupole with the $y(3x^2-y^2)$ component~\cite{doi:10.7566/JPSJ.85.053705,PhysRevB.99.174407}, which results in the $q_y(3q_x^2-q_y^2)$-type magnon band deformation, as shown in Fig.~\ref{fig1}(c)~\cite{PhysRevB.98.165110}. 
This antisymmetric functional form implies that a directional nonreciprocity is coupled with the quadrupole degrees of freedom  (second order of $H$) when dividing $q_y(3q_x^2-q_y^2)$ as $2 q_x \times (q_x q_y)+q_y \times (q_x^2-q_y^2)$. 
As the symmetry $q_x q_y$ and $q_x^2-q_y^2$ are the same as $H_x H_y$ and $H_x^2-H_y^2$, the $y(3x^2-y^2)$-type magnetic toroidal octupole gives rise to the coupling as $2 q_x \times (H_x H_y)+q_y \times (H_x^2-H_y^2) \sim q_x \sin(2\phi)+q_y\cos(2\phi)$.
As $q_x$ and $q_y$ correspond to the polar vector, this decomposition expresses the emergence of in-plane magnetic toroidal dipoles $T_x \sim q_x$ and $T_y \sim q_y$ in the canted AFM state, and explains a qualitative behavior of the result in Figs.~\ref{fig2}(c)-(f) and \ref{fig3}(a). 
Such an emergent magnetic toroidal dipole in the canted AFM state is consistent with the symmetry analysis by using the cluster multipole theory~\cite{PhysRevB.99.174407}.
From the viewpoint of model parameters, the symmetric anisotropic exchange interaction is essential, since it breaks continuous spin rotational symmetry.
The threefold symmetric interaction consists of the product of the dipole and quadrupole degrees of freedom on the basis of the microscopic multipole description~\cite{matsumoto2017symmetry, PhysRevB.98.165110}.
Recently, a similar angle-dependent magneto-electric effect observed in Co$_4$Nb$_2$O$_9$~\cite{Khanh_PhysRevB.93.075117, Khanh_PhysRevB.96.094434} is understood from the multipole aspect~\cite{Yanagi_PhysRevB.97.020404,doi:10.7566/JPSJ.88.094704}.

\section{summary}
\label{summary}

To summarize, we have investigated the behavior of nonreciprocal magnon induced by the nearest-neighbor symmetric anisotropic exchange interaction on a honeycomb AFM.
The antisymmetric nature of magnon bands is qualitatively different from that by the DM interaction. 
Moreover, we have found that the nonreciprocal magnon excitations exhibit peculiar angle-dependent responses under the external magnetic field.
We have also clarified that the nonreciprocal dispersions and angle-dependent responses are related with the emergence of odd-parity magnetic toroidal multipoles, which are accompanied by the cluster AFM structure.
As the antisymmetric modulation of the magnon bands becomes larger while increasing the bond-dependent symmetric anisotropic exchange interaction, the superexchange paths favoring the anisotropic interactions rather than the Heisenberg interaction, such as the Kitaev interaction~\cite{PhysRevLett.102.017205}, will enhance nonreciprocal physical phenomena.

Our mechanism of nonreciprocal magnons is expected to be observed in various honeycomb AFMs including the transition-metal tricalcogenide $\rm{MnPS_3}$~\cite{PhysRevB.82.100408, Li3738, PhysRevB.91.235425, PhysRevB.96.134425} and rare-earth metallic compound $\rm{ErNi_3Ga_9}$~\cite{e1ddc90bb22e4448a040d6ce3fdca500}, where the $z$-AFM state becomes the ground state.
In these materials, the nonreciprocal magnon excitations will be observed in both microscopic and macroscopic experiments.
Microscopically, the nonreciprocal magnon spectra can be detected by the inelastic neutron scattering experiment.
On the other hand, from a macroscopic viewpoint, the angle-dependent nonreciprocal magneto-optical and nonreciprocal spin Seebeck effects can be observed under an in-plane magnetic field.
As the nature of the asymmetric deformation of the magnon band is qualitatively different from that by the DM interaction, our mechanism will provide a deep understanding of further nonreciprocal magnon physics.

\begin{acknowledgments}
We would like to thank T. J. Sato for fruitful discussions.
This research was supported by JSPS KAKENHI Grants Numbers JP18H04296 (J-Physics), JP18K13488, JP19K03752, and JP19H01834. 
This work was also supported by the Toyota Riken Scholarship.
Parts of the numerical calculations were performed in the supercomputing systems in ISSP, the University of Tokyo.
\end{acknowledgments}

\appendix

\section{Ground-state spin configuration}

\label{GS}
In this Appendix, we show that the collinear AFM orderings with the $z$ spin component are stabilized for $H=0$ and the canted AFM ordering are stabilized for $\mathbf{H}=H(\cos{\phi}, \sin{\phi},0)$ by assuming the two-sublattice ordering and $J^z>J>0$. 
We use four variational parameters $(\theta_{\rm A},\phi_{\rm A}, \theta_{\rm B},\phi_{\rm B})$ representing the $\eta(={\rm A}$ and B)-sublattice spin state: $\mathbf{S}_{\eta}=S(\sin{\theta_{\eta}}\cos{\phi_{\eta}},\sin{\theta_{\eta}}\sin{\phi_{\eta}},\cos{\theta_{\eta}})$. 
Then, the spin Hamiltonian in Eq.~\eqref{eq1} is rewritten as
\begin{align}
\label{A1}
\mathcal{H}
=&
\frac{3NS^2}{2}[J\sin{\theta_{\rm{A}}}\sin{\theta_{\rm{B}}}\cos(\phi_{\rm{A}}-\phi_{\rm{B}})
+J^z\cos{\theta_{\rm{A}}}\cos{\theta_{\rm{B}}}],
\end{align}
where $N$ is the number of total spins.
For $J^z>J>0$, the staggered collinear AFM order is stabilized to satisfy $\theta_{\rm{A}}=\theta$, $\theta_{\rm{B}}=\pi-\theta$, $\phi_{\rm{A}}=\phi$, and $\phi_{\rm{B}}=\pi+\phi$.
Then, Eq.~\eqref{A1} reduces to
\begin{align}
\label{A2}
\mathcal{H}=&-\frac{3NS^2}{2}[(J^z-J)\cos^2{\theta}+J].
\end{align} 
Thus, the ground state is realized at $ \theta=0$ or $\pi$ when $J^z > J$.

Next, we show the optimal spin configuration under $\mathbf{H}$.
To gain the Zeeman energy, the staggered AFM moment along the $z$ direction is canted along the magnetic-field direction, whose spin ansatz is represented by $\mathbf{S}_{i{\rm A}}=S(\sin{\theta}\cos{\phi}, \sin{\theta}\sin{\phi}, \cos{\theta})$ and $\mathbf{S}_{i{\rm B}}=S(\sin{\theta}\cos{\phi}, \sin{\theta}\sin{\phi}, -\cos{\theta})$.
Then, the spin Hamiltonian in Eq.~\eqref{eq1} is rewritten as
\begin{align}
\label{A3}
\mathcal{H}
=\frac{N}{2}\Bigg\{&3S^2(J^z+J)\left[\sin{\theta}-\frac{H}{3S(J^z+J)}\right]^2\nonumber\\
&-3S^2J^z-\frac{H^2}{3(J^z+J)}\Bigg\}.
\end{align}
By minimizing the ground state energy of the model in Eq.~\eqref{A3} with respect to $\theta$, the optimal canted value of $\theta$ is obtained by $\theta=\sin^{-1}\left[H/3S(J+J^z)\right]$.
The spin waves in Sec.~\ref{result} are calculated for the obtained spin configurations.

\section{Nonreciprocal magnons in the presence of the Dzyaloshinskii-Moriya interaction
\label{DM}}

\begin{figure}[h]
\includegraphics[width=1.0\linewidth]{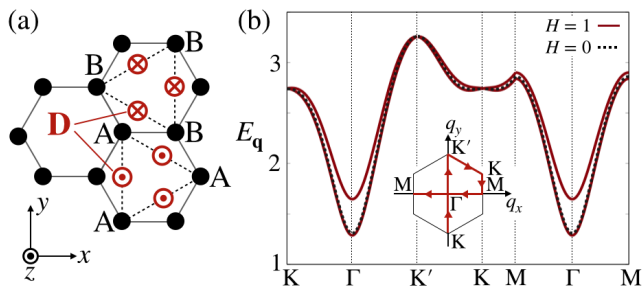}
\caption{\label{fig4}  (a) The DM vector in the honeycomb structure. 
(b) The magnon dispersions in the model in Eq.~\eqref{B1} at $(J, J^z, D)=(0.9, 1, 0.05)$.
The red solid (black dotted) lines represent the result at $H=1$ ($H=0$).
In the inset, the first Brillouin zone is shown. }
\end{figure}

In this Appendix, we discuss nonreciprocal magnons induced by the DM interaction for comparison.
The localized spin model including the next-nearest-neighbor interaction in the honeycomb structure is given by
\begin{align}
\label{B1}
\mathcal{H}=&
\sum_{\langle ij \rangle} 
 \Big[
 J(S^+_{i{\rm{A}}}S^-_{j{\rm{B}}}+S^-_{i{\rm{A}}}S^+_{j{\rm{B}}}) +J^z S^z_{i{\rm{A}}} S^z_{j{\rm{B}}} \Big] \nonumber \\
&+\sum_{i, j}D\left(\mathbf{S}_{ i {\rm{A}}}\times \mathbf{S}_{j {\rm{A}}}-\mathbf{S}_{i {\rm{B}}}\times \mathbf{S}_{j {\rm{B}}}\right)^z
 -\sum_{i, \eta}
 \mathbf{H}\cdot \mathbf{S}_{i \eta}.
\end{align}
We here take into account the DM interaction appearing in the next-nearest-neighbor spins instead of $J^a$ in the model in Eq.~\eqref{eq1}, as shown in Fig~\ref{fig4}(a).
The collinear AFM ordering along the $z$ direction is stabilized in the model in Eq.~\eqref{B1} by choosing the parameter $J^z>J \gg D$.

Figure~\ref{fig4}(b) shows the magnon dispersion for nonzero $D=0.05$ at $H=1$ (red solid line) and $H=0$ (black dotted line).
The black dotted line shows magnon-band inclination on K-$\Gamma$-$\rm{K}'$ line by DM interaction.
The red solid line shows the result in the presence of the in-plane magnetic field independent of the field direction.
This tendency is clearly different from that by the symmetric anisotropic exchange interaction, which gives rise to direction-dependent nonreciprocal dispersions, in Sec.~\ref{Hneq0} as well as that in the other nonreciprocal systems mentioned in the introduction~\cite{cortes2013influence, zhang2015plane, cho2015thickness, tacchi2017interfacial, chaurasiya2018dependence,grunberg1986layered, zhang1987spin, di2015enhancement, gallardo2019reconfigurable, albisetti2020optically}.

\bibliography{ref}

\begin{thebibliography}{65}%
\makeatletter
\providecommand \@ifxundefined [1]{%
 \@ifx{#1\undefined}
}%
\providecommand \@ifnum [1]{%
 \ifnum #1\expandafter \@firstoftwo
 \else \expandafter \@secondoftwo
 \fi
}%
\providecommand \@ifx [1]{%
 \ifx #1\expandafter \@firstoftwo
 \else \expandafter \@secondoftwo
 \fi
}%
\providecommand \natexlab [1]{#1}%
\providecommand \enquote  [1]{``#1''}%
\providecommand \bibnamefont  [1]{#1}%
\providecommand \bibfnamefont [1]{#1}%
\providecommand \citenamefont [1]{#1}%
\providecommand \href@noop [0]{\@secondoftwo}%
\providecommand \href [0]{\begingroup \@sanitize@url \@href}%
\providecommand \@href[1]{\@@startlink{#1}\@@href}%
\providecommand \@@href[1]{\endgroup#1\@@endlink}%
\providecommand \@sanitize@url [0]{\catcode `\\12\catcode `\$12\catcode
  `\&12\catcode `\#12\catcode `\^12\catcode `\_12\catcode `\%12\relax}%
\providecommand \@@startlink[1]{}%
\providecommand \@@endlink[0]{}%
\providecommand \url  [0]{\begingroup\@sanitize@url \@url }%
\providecommand \@url [1]{\endgroup\@href {#1}{\urlprefix }}%
\providecommand \urlprefix  [0]{URL }%
\providecommand \Eprint [0]{\href }%
\providecommand \doibase [0]{https://doi.org/}%
\providecommand \selectlanguage [0]{\@gobble}%
\providecommand \bibinfo  [0]{\@secondoftwo}%
\providecommand \bibfield  [0]{\@secondoftwo}%
\providecommand \translation [1]{[#1]}%
\providecommand \BibitemOpen [0]{}%
\providecommand \bibitemStop [0]{}%
\providecommand \bibitemNoStop [0]{.\EOS\space}%
\providecommand \EOS [0]{\spacefactor3000\relax}%
\providecommand \BibitemShut  [1]{\csname bibitem#1\endcsname}%
\let\auto@bib@innerbib\@empty
\bibitem [{\citenamefont {{Curie, P.}}(1894)}]{Curie}%
  \BibitemOpen
  \bibfield  {author} {\bibinfo {author} {\bibnamefont {{Curie, P.}}},\
  }\bibfield  {title} {\bibinfo {title} {Sur la sym\'etrie dans les
  ph\'enom\`enes physiques, sym\'etrie d'un champ \'electrique et d'un champ
  magn\'etique},\ }\href {https://doi.org/10.1051/jphystap:018940030039300}
  {\bibfield  {journal} {\bibinfo  {journal} {J. Phys. Theor. Appl.}\ }\textbf
  {\bibinfo {volume} {3}},\ \bibinfo {pages} {393} (\bibinfo {year}
  {1894})}\BibitemShut {NoStop}%
\bibitem [{\citenamefont {Fiebig}(2005)}]{Fiebig_2005}%
  \BibitemOpen
  \bibfield  {author} {\bibinfo {author} {\bibfnamefont {M.}~\bibnamefont
  {Fiebig}},\ }\bibfield  {title} {\bibinfo {title} {Revival of the
  magnetoelectric effect},\ }\href {https://doi.org/10.1088/0022-3727/38/8/r01}
  {\bibfield  {journal} {\bibinfo  {journal} {J. Phys. D: Appl. Phys.}\
  }\textbf {\bibinfo {volume} {38}},\ \bibinfo {pages} {R123} (\bibinfo {year}
  {2005})}\BibitemShut {NoStop}%
\bibitem [{\citenamefont {Khomskii}(2009)}]{khomskii2009trend}%
  \BibitemOpen
  \bibfield  {author} {\bibinfo {author} {\bibfnamefont {D.}~\bibnamefont
  {Khomskii}},\ }\bibfield  {title} {\bibinfo {title} {Trend: Classifying
  multiferroics: Mechanisms and effects},\ }\href@noop {} {\bibfield  {journal}
  {\bibinfo  {journal} {Physics}\ }\textbf {\bibinfo {volume} {2}},\ \bibinfo
  {pages} {20} (\bibinfo {year} {2009})}\BibitemShut {NoStop}%
\bibitem [{\citenamefont {Wakatsuki}\ \emph {et~al.}(2017)\citenamefont
  {Wakatsuki}, \citenamefont {Saito}, \citenamefont {Hoshino}, \citenamefont
  {Itahashi}, \citenamefont {Ideue}, \citenamefont {Ezawa}, \citenamefont
  {Iwasa},\ and\ \citenamefont {Nagaosa}}]{wakatsuki2017nonreciprocal}%
  \BibitemOpen
  \bibfield  {author} {\bibinfo {author} {\bibfnamefont {R.}~\bibnamefont
  {Wakatsuki}}, \bibinfo {author} {\bibfnamefont {Y.}~\bibnamefont {Saito}},
  \bibinfo {author} {\bibfnamefont {S.}~\bibnamefont {Hoshino}}, \bibinfo
  {author} {\bibfnamefont {Y.~M.}\ \bibnamefont {Itahashi}}, \bibinfo {author}
  {\bibfnamefont {T.}~\bibnamefont {Ideue}}, \bibinfo {author} {\bibfnamefont
  {M.}~\bibnamefont {Ezawa}}, \bibinfo {author} {\bibfnamefont
  {Y.}~\bibnamefont {Iwasa}},\ and\ \bibinfo {author} {\bibfnamefont
  {N.}~\bibnamefont {Nagaosa}},\ }\bibfield  {title} {\bibinfo {title}
  {Nonreciprocal charge transport in noncentrosymmetric superconductors},\
  }\href@noop {} {\bibfield  {journal} {\bibinfo  {journal} {Sci. Adv.}\
  }\textbf {\bibinfo {volume} {3}},\ \bibinfo {pages} {e1602390} (\bibinfo
  {year} {2017})}\BibitemShut {NoStop}%
\bibitem [{\citenamefont {Mochizuki}\ and\ \citenamefont
  {Seki}(2013)}]{PhysRevB.87.134403}%
  \BibitemOpen
  \bibfield  {author} {\bibinfo {author} {\bibfnamefont {M.}~\bibnamefont
  {Mochizuki}}\ and\ \bibinfo {author} {\bibfnamefont {S.}~\bibnamefont
  {Seki}},\ }\bibfield  {title} {\bibinfo {title} {Magnetoelectric resonances
  and predicted microwave diode effect of the skyrmion crystal in a
  multiferroic chiral-lattice magnet},\ }\href
  {https://doi.org/10.1103/PhysRevB.87.134403} {\bibfield  {journal} {\bibinfo
  {journal} {Phys. Rev. B}\ }\textbf {\bibinfo {volume} {87}},\ \bibinfo
  {pages} {134403} (\bibinfo {year} {2013})}\BibitemShut {NoStop}%
\bibitem [{\citenamefont {Saito}\ \emph {et~al.}(2018)\citenamefont {Saito},
  \citenamefont {Uenishi}, \citenamefont {Miura}, \citenamefont {Tabata},
  \citenamefont {Hidaka}, \citenamefont {Yanagisawa},\ and\ \citenamefont
  {Amitsuka}}]{doi:10.7566/JPSJ.87.033702}%
  \BibitemOpen
  \bibfield  {author} {\bibinfo {author} {\bibfnamefont {H.}~\bibnamefont
  {Saito}}, \bibinfo {author} {\bibfnamefont {K.}~\bibnamefont {Uenishi}},
  \bibinfo {author} {\bibfnamefont {N.}~\bibnamefont {Miura}}, \bibinfo
  {author} {\bibfnamefont {C.}~\bibnamefont {Tabata}}, \bibinfo {author}
  {\bibfnamefont {H.}~\bibnamefont {Hidaka}}, \bibinfo {author} {\bibfnamefont
  {T.}~\bibnamefont {Yanagisawa}},\ and\ \bibinfo {author} {\bibfnamefont
  {H.}~\bibnamefont {Amitsuka}},\ }\bibfield  {title} {\bibinfo {title}
  {Evidence of a new current-induced magnetoelectric effect in a toroidal
  magnetic ordered state of $\rm{UNi_4B}$},\ }\href
  {https://doi.org/10.7566/JPSJ.87.033702} {\bibfield  {journal} {\bibinfo
  {journal} {J. Phys. Soc. Jpn.}\ }\textbf {\bibinfo {volume} {87}},\ \bibinfo
  {pages} {033702} (\bibinfo {year} {2018})}\BibitemShut {NoStop}%
\bibitem [{\citenamefont {Shiomi}\ \emph {et~al.}(2018)\citenamefont {Shiomi},
  \citenamefont {Akiba}, \citenamefont {Takahashi},\ and\ \citenamefont
  {Ishiwata}}]{magpiezo1}%
  \BibitemOpen
  \bibfield  {author} {\bibinfo {author} {\bibfnamefont {Y.}~\bibnamefont
  {Shiomi}}, \bibinfo {author} {\bibfnamefont {T.}~\bibnamefont {Akiba}},
  \bibinfo {author} {\bibfnamefont {H.}~\bibnamefont {Takahashi}},\ and\
  \bibinfo {author} {\bibfnamefont {S.}~\bibnamefont {Ishiwata}},\ }\bibfield
  {title} {\bibinfo {title} {Giant piezoelectric response in superionic polar
  semiconductor},\ }\href {https://doi.org/10.1002/aelm.201800174} {\bibfield
  {journal} {\bibinfo  {journal} {Adv. Electron. Mater.}\ }\textbf {\bibinfo
  {volume} {4}} (\bibinfo {year} {2018})}\BibitemShut {NoStop}%
\bibitem [{\citenamefont {Shiomi}\ \emph {et~al.}(2019)\citenamefont {Shiomi},
  \citenamefont {Watanabe}, \citenamefont {Masuda}, \citenamefont {Takahashi},
  \citenamefont {Yanase},\ and\ \citenamefont
  {Ishiwata}}]{PhysRevLett.122.127207}%
  \BibitemOpen
  \bibfield  {author} {\bibinfo {author} {\bibfnamefont {Y.}~\bibnamefont
  {Shiomi}}, \bibinfo {author} {\bibfnamefont {H.}~\bibnamefont {Watanabe}},
  \bibinfo {author} {\bibfnamefont {H.}~\bibnamefont {Masuda}}, \bibinfo
  {author} {\bibfnamefont {H.}~\bibnamefont {Takahashi}}, \bibinfo {author}
  {\bibfnamefont {Y.}~\bibnamefont {Yanase}},\ and\ \bibinfo {author}
  {\bibfnamefont {S.}~\bibnamefont {Ishiwata}},\ }\bibfield  {title} {\bibinfo
  {title} {Observation of a magnetopiezoelectric effect in the
  antiferromagnetic metal $\rm{Eu Mn Bi_2}$},\ }\href
  {https://doi.org/10.1103/PhysRevLett.122.127207} {\bibfield  {journal}
  {\bibinfo  {journal} {Phys. Rev. Lett.}\ }\textbf {\bibinfo {volume} {122}},\
  \bibinfo {pages} {127207} (\bibinfo {year} {2019})}\BibitemShut {NoStop}%
\bibitem [{\citenamefont {Melcher}(1973)}]{PhysRevLett.30.125}%
  \BibitemOpen
  \bibfield  {author} {\bibinfo {author} {\bibfnamefont {R.~L.}\ \bibnamefont
  {Melcher}},\ }\bibfield  {title} {\bibinfo {title} {Linear contribution to
  spatial dispersion in the spin-wave spectrum of ferromagnets},\ }\href
  {https://doi.org/10.1103/PhysRevLett.30.125} {\bibfield  {journal} {\bibinfo
  {journal} {Phys. Rev. Lett.}\ }\textbf {\bibinfo {volume} {30}},\ \bibinfo
  {pages} {125} (\bibinfo {year} {1973})}\BibitemShut {NoStop}%
\bibitem [{\citenamefont {Kataoka}(1987)}]{doi:10.1143/JPSJ.56.3635}%
  \BibitemOpen
  \bibfield  {author} {\bibinfo {author} {\bibfnamefont {M.}~\bibnamefont
  {Kataoka}},\ }\bibfield  {title} {\bibinfo {title} {Spin waves in systems
  with long period helical spin density waves due to the antisymmetric and
  symmetric exchange interactions},\ }\href
  {https://doi.org/10.1143/JPSJ.56.3635} {\bibfield  {journal} {\bibinfo
  {journal} {J. Phys. Soc. Jpn.}\ }\textbf {\bibinfo {volume} {56}},\ \bibinfo
  {pages} {3635} (\bibinfo {year} {1987})}\BibitemShut {NoStop}%
\bibitem [{\citenamefont {Cort{\'e}s-Ortu{\~n}o}\ and\ \citenamefont
  {Landeros}(2013)}]{cortes2013influence}%
  \BibitemOpen
  \bibfield  {author} {\bibinfo {author} {\bibfnamefont {D.}~\bibnamefont
  {Cort{\'e}s-Ortu{\~n}o}}\ and\ \bibinfo {author} {\bibfnamefont
  {P.}~\bibnamefont {Landeros}},\ }\bibfield  {title} {\bibinfo {title}
  {Influence of the $\rm{Dzyaloshinskii}$-$\rm{Moriya}$ interaction on the
  spin-wave spectra of thin films},\ }\href@noop {} {\bibfield  {journal}
  {\bibinfo  {journal} {J. Phys.: Condens. Matter}\ }\textbf {\bibinfo {volume}
  {25}},\ \bibinfo {pages} {156001} (\bibinfo {year} {2013})}\BibitemShut
  {NoStop}%
\bibitem [{\citenamefont {Moon}\ \emph {et~al.}(2013)\citenamefont {Moon},
  \citenamefont {Seo}, \citenamefont {Lee}, \citenamefont {Kim}, \citenamefont
  {Ryu}, \citenamefont {Lee}, \citenamefont {McMichael},\ and\ \citenamefont
  {Stiles}}]{PhysRevB.88.184404}%
  \BibitemOpen
  \bibfield  {author} {\bibinfo {author} {\bibfnamefont {J.-H.}\ \bibnamefont
  {Moon}}, \bibinfo {author} {\bibfnamefont {S.-M.}\ \bibnamefont {Seo}},
  \bibinfo {author} {\bibfnamefont {K.-J.}\ \bibnamefont {Lee}}, \bibinfo
  {author} {\bibfnamefont {K.-W.}\ \bibnamefont {Kim}}, \bibinfo {author}
  {\bibfnamefont {J.}~\bibnamefont {Ryu}}, \bibinfo {author} {\bibfnamefont
  {H.-W.}\ \bibnamefont {Lee}}, \bibinfo {author} {\bibfnamefont {R.~D.}\
  \bibnamefont {McMichael}},\ and\ \bibinfo {author} {\bibfnamefont {M.~D.}\
  \bibnamefont {Stiles}},\ }\bibfield  {title} {\bibinfo {title} {Spin-wave
  propagation in the presence of interfacial
  $\rm{Dzyaloshinskii}$-$\rm{Moriya}$ interaction},\ }\href
  {https://doi.org/10.1103/PhysRevB.88.184404} {\bibfield  {journal} {\bibinfo
  {journal} {Phys. Rev. B}\ }\textbf {\bibinfo {volume} {88}},\ \bibinfo
  {pages} {184404} (\bibinfo {year} {2013})}\BibitemShut {NoStop}%
\bibitem [{\citenamefont {Hayami}\ \emph {et~al.}(2016)\citenamefont {Hayami},
  \citenamefont {Kusunose},\ and\ \citenamefont
  {Motome}}]{doi:10.7566/JPSJ.85.053705}%
  \BibitemOpen
  \bibfield  {author} {\bibinfo {author} {\bibfnamefont {S.}~\bibnamefont
  {Hayami}}, \bibinfo {author} {\bibfnamefont {H.}~\bibnamefont {Kusunose}},\
  and\ \bibinfo {author} {\bibfnamefont {Y.}~\bibnamefont {Motome}},\
  }\bibfield  {title} {\bibinfo {title} {Asymmetric magnon excitation by
  spontaneous toroidal ordering},\ }\href
  {https://doi.org/10.7566/JPSJ.85.053705} {\bibfield  {journal} {\bibinfo
  {journal} {J. Phys. Soc. Jpn.}\ }\textbf {\bibinfo {volume} {85}},\ \bibinfo
  {pages} {053705} (\bibinfo {year} {2016})}\BibitemShut {NoStop}%
\bibitem [{\citenamefont {Ghader}\ and\ \citenamefont
  {Khater}(2019)}]{10.1038/s41598-019-51646-3}%
  \BibitemOpen
  \bibfield  {author} {\bibinfo {author} {\bibfnamefont {D.}~\bibnamefont
  {Ghader}}\ and\ \bibinfo {author} {\bibfnamefont {A.}~\bibnamefont
  {Khater}},\ }\bibfield  {title} {\bibinfo {title} {A new class of
  nonreciprocal spin waves on the edges of 2d antiferromagnetic honeycomb
  nanoribbons},\ }\href {https://doi.org/10.7566/JPSJ.88.094704} {\bibfield
  {journal} {\bibinfo  {journal} {Sci. Rep.}\ }\textbf {\bibinfo {volume}
  {9}},\ \bibinfo {pages} {2045} (\bibinfo {year} {2019})}\BibitemShut
  {NoStop}%
\bibitem [{\citenamefont {Kawano}\ \emph {et~al.}(2019)\citenamefont {Kawano},
  \citenamefont {Onose},\ and\ \citenamefont {Hotta}}]{kawano2019designing}%
  \BibitemOpen
  \bibfield  {author} {\bibinfo {author} {\bibfnamefont {M.}~\bibnamefont
  {Kawano}}, \bibinfo {author} {\bibfnamefont {Y.}~\bibnamefont {Onose}},\ and\
  \bibinfo {author} {\bibfnamefont {C.}~\bibnamefont {Hotta}},\ }\bibfield
  {title} {\bibinfo {title} {Designing $\rm{Rashba}$--$\rm{Dresselhaus}$ effect
  in magnetic insulators},\ }\href@noop {} {\bibfield  {journal} {\bibinfo
  {journal} {Commun. Phys.}\ }\textbf {\bibinfo {volume} {2}},\ \bibinfo
  {pages} {27} (\bibinfo {year} {2019})}\BibitemShut {NoStop}%
\bibitem [{\citenamefont {Kawano}\ and\ \citenamefont
  {Hotta}(2019)}]{PhysRevB.100.174402}%
  \BibitemOpen
  \bibfield  {author} {\bibinfo {author} {\bibfnamefont {M.}~\bibnamefont
  {Kawano}}\ and\ \bibinfo {author} {\bibfnamefont {C.}~\bibnamefont {Hotta}},\
  }\bibfield  {title} {\bibinfo {title} {Discovering momentum-dependent magnon
  spin texture in insulating antiferromagnets: Role of the $\rm{Kitaev}$
  interaction},\ }\href {https://doi.org/10.1103/PhysRevB.100.174402}
  {\bibfield  {journal} {\bibinfo  {journal} {Phys. Rev. B}\ }\textbf {\bibinfo
  {volume} {100}},\ \bibinfo {pages} {174402} (\bibinfo {year}
  {2019})}\BibitemShut {NoStop}%
\bibitem [{\citenamefont {Dzyaloshinsky}(1958)}]{DZYALOSHINSKY1958241}%
  \BibitemOpen
  \bibfield  {author} {\bibinfo {author} {\bibfnamefont {I.}~\bibnamefont
  {Dzyaloshinsky}},\ }\bibfield  {title} {\bibinfo {title} {A thermodynamic
  theory of “weak” ferromagnetism of antiferromagnetics},\ }\href
  {https://doi.org/https://doi.org/10.1016/0022-3697(58)90076-3} {\bibfield
  {journal} {\bibinfo  {journal} {J. Phys. Chem. Solids}\ }\textbf {\bibinfo
  {volume} {4}},\ \bibinfo {pages} {241 } (\bibinfo {year} {1958})}\BibitemShut
  {NoStop}%
\bibitem [{\citenamefont {Moriya}(1960)}]{PhysRev.120.91}%
  \BibitemOpen
  \bibfield  {author} {\bibinfo {author} {\bibfnamefont {T.}~\bibnamefont
  {Moriya}},\ }\bibfield  {title} {\bibinfo {title} {Anisotropic superexchange
  interaction and weak ferromagnetism},\ }\href
  {https://doi.org/10.1103/PhysRev.120.91} {\bibfield  {journal} {\bibinfo
  {journal} {Phys. Rev.}\ }\textbf {\bibinfo {volume} {120}},\ \bibinfo {pages}
  {91} (\bibinfo {year} {1960})}\BibitemShut {NoStop}%
\bibitem [{\citenamefont {Iguchi}\ \emph {et~al.}(2015)\citenamefont {Iguchi},
  \citenamefont {Uemura}, \citenamefont {Ueno},\ and\ \citenamefont
  {Onose}}]{PhysRevB.92.184419}%
  \BibitemOpen
  \bibfield  {author} {\bibinfo {author} {\bibfnamefont {Y.}~\bibnamefont
  {Iguchi}}, \bibinfo {author} {\bibfnamefont {S.}~\bibnamefont {Uemura}},
  \bibinfo {author} {\bibfnamefont {K.}~\bibnamefont {Ueno}},\ and\ \bibinfo
  {author} {\bibfnamefont {Y.}~\bibnamefont {Onose}},\ }\bibfield  {title}
  {\bibinfo {title} {Nonreciprocal magnon propagation in a noncentrosymmetric
  ferromagnet $\rm{LiFe_5O_8}$},\ }\href
  {https://doi.org/10.1103/PhysRevB.92.184419} {\bibfield  {journal} {\bibinfo
  {journal} {Phys. Rev. B}\ }\textbf {\bibinfo {volume} {92}},\ \bibinfo
  {pages} {184419} (\bibinfo {year} {2015})}\BibitemShut {NoStop}%
\bibitem [{\citenamefont {Zhang}\ \emph {et~al.}(2015)\citenamefont {Zhang},
  \citenamefont {Di}, \citenamefont {Lim}, \citenamefont {Ng}, \citenamefont
  {Kuok}, \citenamefont {Yu}, \citenamefont {Yoon}, \citenamefont {Qiu},\ and\
  \citenamefont {Yang}}]{zhang2015plane}%
  \BibitemOpen
  \bibfield  {author} {\bibinfo {author} {\bibfnamefont {V.~L.}\ \bibnamefont
  {Zhang}}, \bibinfo {author} {\bibfnamefont {K.}~\bibnamefont {Di}}, \bibinfo
  {author} {\bibfnamefont {H.~S.}\ \bibnamefont {Lim}}, \bibinfo {author}
  {\bibfnamefont {S.~C.}\ \bibnamefont {Ng}}, \bibinfo {author} {\bibfnamefont
  {M.~H.}\ \bibnamefont {Kuok}}, \bibinfo {author} {\bibfnamefont
  {J.}~\bibnamefont {Yu}}, \bibinfo {author} {\bibfnamefont {J.}~\bibnamefont
  {Yoon}}, \bibinfo {author} {\bibfnamefont {X.}~\bibnamefont {Qiu}},\ and\
  \bibinfo {author} {\bibfnamefont {H.}~\bibnamefont {Yang}},\ }\bibfield
  {title} {\bibinfo {title} {In-plane angular dependence of the spin-wave
  nonreciprocity of an ultrathin film with $\rm{Dzyaloshinskii}$-$\rm{Moriya}$
  interaction},\ }\href@noop {} {\bibfield  {journal} {\bibinfo  {journal}
  {Appl. Phys. Lett.}\ }\textbf {\bibinfo {volume} {107}},\ \bibinfo {pages}
  {022402} (\bibinfo {year} {2015})}\BibitemShut {NoStop}%
\bibitem [{\citenamefont {Cho}\ \emph {et~al.}(2015)\citenamefont {Cho},
  \citenamefont {Kim}, \citenamefont {Lee}, \citenamefont {Kim}, \citenamefont
  {Lavrijsen}, \citenamefont {Solignac}, \citenamefont {Yin}, \citenamefont
  {Han}, \citenamefont {Van~Hoof}, \citenamefont {Swagten} \emph
  {et~al.}}]{cho2015thickness}%
  \BibitemOpen
  \bibfield  {author} {\bibinfo {author} {\bibfnamefont {J.}~\bibnamefont
  {Cho}}, \bibinfo {author} {\bibfnamefont {N.-H.}\ \bibnamefont {Kim}},
  \bibinfo {author} {\bibfnamefont {S.}~\bibnamefont {Lee}}, \bibinfo {author}
  {\bibfnamefont {J.-S.}\ \bibnamefont {Kim}}, \bibinfo {author} {\bibfnamefont
  {R.}~\bibnamefont {Lavrijsen}}, \bibinfo {author} {\bibfnamefont
  {A.}~\bibnamefont {Solignac}}, \bibinfo {author} {\bibfnamefont
  {Y.}~\bibnamefont {Yin}}, \bibinfo {author} {\bibfnamefont {D.-S.}\
  \bibnamefont {Han}}, \bibinfo {author} {\bibfnamefont {N.~J.}\ \bibnamefont
  {Van~Hoof}}, \bibinfo {author} {\bibfnamefont {H.~J.}\ \bibnamefont
  {Swagten}}, \emph {et~al.},\ }\bibfield  {title} {\bibinfo {title} {Thickness
  dependence of the interfacial $\rm{Dzyaloshinskii}$-$\rm{Moriya}$ interaction
  in inversion symmetry broken systems},\ }\href@noop {} {\bibfield  {journal}
  {\bibinfo  {journal} {Nat. Commun.}\ }\textbf {\bibinfo {volume} {6}},\
  \bibinfo {pages} {1} (\bibinfo {year} {2015})}\BibitemShut {NoStop}%
\bibitem [{\citenamefont {Sato}\ \emph {et~al.}(2016)\citenamefont {Sato},
  \citenamefont {Okuyama}, \citenamefont {Hong}, \citenamefont {Kikkawa},
  \citenamefont {Taguchi}, \citenamefont {Arima},\ and\ \citenamefont
  {Tokura}}]{PhysRevB.94.144420}%
  \BibitemOpen
  \bibfield  {author} {\bibinfo {author} {\bibfnamefont {T.~J.}\ \bibnamefont
  {Sato}}, \bibinfo {author} {\bibfnamefont {D.}~\bibnamefont {Okuyama}},
  \bibinfo {author} {\bibfnamefont {T.}~\bibnamefont {Hong}}, \bibinfo {author}
  {\bibfnamefont {A.}~\bibnamefont {Kikkawa}}, \bibinfo {author} {\bibfnamefont
  {Y.}~\bibnamefont {Taguchi}}, \bibinfo {author} {\bibfnamefont {T.-h.}\
  \bibnamefont {Arima}},\ and\ \bibinfo {author} {\bibfnamefont
  {Y.}~\bibnamefont {Tokura}},\ }\bibfield  {title} {\bibinfo {title} {Magnon
  dispersion shift in the induced ferromagnetic phase of noncentrosymmetric
  $\rm{MnSi}$},\ }\href {https://doi.org/10.1103/PhysRevB.94.144420} {\bibfield
   {journal} {\bibinfo  {journal} {Phys. Rev. B}\ }\textbf {\bibinfo {volume}
  {94}},\ \bibinfo {pages} {144420} (\bibinfo {year} {2016})}\BibitemShut
  {NoStop}%
\bibitem [{\citenamefont {Seki}\ \emph {et~al.}(2016)\citenamefont {Seki},
  \citenamefont {Okamura}, \citenamefont {Kondou}, \citenamefont {Shibata},
  \citenamefont {Kubota}, \citenamefont {Takagi}, \citenamefont {Kagawa},
  \citenamefont {Kawasaki}, \citenamefont {Tatara}, \citenamefont {Otani},\
  and\ \citenamefont {Tokura}}]{PhysRevB.93.235131}%
  \BibitemOpen
  \bibfield  {author} {\bibinfo {author} {\bibfnamefont {S.}~\bibnamefont
  {Seki}}, \bibinfo {author} {\bibfnamefont {Y.}~\bibnamefont {Okamura}},
  \bibinfo {author} {\bibfnamefont {K.}~\bibnamefont {Kondou}}, \bibinfo
  {author} {\bibfnamefont {K.}~\bibnamefont {Shibata}}, \bibinfo {author}
  {\bibfnamefont {M.}~\bibnamefont {Kubota}}, \bibinfo {author} {\bibfnamefont
  {R.}~\bibnamefont {Takagi}}, \bibinfo {author} {\bibfnamefont
  {F.}~\bibnamefont {Kagawa}}, \bibinfo {author} {\bibfnamefont
  {M.}~\bibnamefont {Kawasaki}}, \bibinfo {author} {\bibfnamefont
  {G.}~\bibnamefont {Tatara}}, \bibinfo {author} {\bibfnamefont
  {Y.}~\bibnamefont {Otani}},\ and\ \bibinfo {author} {\bibfnamefont
  {Y.}~\bibnamefont {Tokura}},\ }\bibfield  {title} {\bibinfo {title}
  {Magnetochiral nonreciprocity of volume spin wave propagation in
  chiral-lattice ferromagnets},\ }\href
  {https://doi.org/10.1103/PhysRevB.93.235131} {\bibfield  {journal} {\bibinfo
  {journal} {Phys. Rev. B}\ }\textbf {\bibinfo {volume} {93}},\ \bibinfo
  {pages} {235131} (\bibinfo {year} {2016})}\BibitemShut {NoStop}%
\bibitem [{\citenamefont {Gitgeatpong}\ \emph {et~al.}(2017)\citenamefont
  {Gitgeatpong}, \citenamefont {Zhao}, \citenamefont {Piyawongwatthana},
  \citenamefont {Qiu}, \citenamefont {Harriger}, \citenamefont {Butch},
  \citenamefont {Sato},\ and\ \citenamefont {Matan}}]{PhysRevLett.119.047201}%
  \BibitemOpen
  \bibfield  {author} {\bibinfo {author} {\bibfnamefont {G.}~\bibnamefont
  {Gitgeatpong}}, \bibinfo {author} {\bibfnamefont {Y.}~\bibnamefont {Zhao}},
  \bibinfo {author} {\bibfnamefont {P.}~\bibnamefont {Piyawongwatthana}},
  \bibinfo {author} {\bibfnamefont {Y.}~\bibnamefont {Qiu}}, \bibinfo {author}
  {\bibfnamefont {L.~W.}\ \bibnamefont {Harriger}}, \bibinfo {author}
  {\bibfnamefont {N.~P.}\ \bibnamefont {Butch}}, \bibinfo {author}
  {\bibfnamefont {T.~J.}\ \bibnamefont {Sato}},\ and\ \bibinfo {author}
  {\bibfnamefont {K.}~\bibnamefont {Matan}},\ }\bibfield  {title} {\bibinfo
  {title} {Nonreciprocal magnons and symmetry-breaking in the
  noncentrosymmetric antiferromagnet},\ }\href
  {https://doi.org/10.1103/PhysRevLett.119.047201} {\bibfield  {journal}
  {\bibinfo  {journal} {Phys. Rev. Lett.}\ }\textbf {\bibinfo {volume} {119}},\
  \bibinfo {pages} {047201} (\bibinfo {year} {2017})}\BibitemShut {NoStop}%
\bibitem [{\citenamefont {Takagi}\ \emph {et~al.}(2017)\citenamefont {Takagi},
  \citenamefont {Morikawa}, \citenamefont {Karube}, \citenamefont {Kanazawa},
  \citenamefont {Shibata}, \citenamefont {Tatara}, \citenamefont {Tokunaga},
  \citenamefont {Arima}, \citenamefont {Taguchi}, \citenamefont {Tokura},\ and\
  \citenamefont {Seki}}]{PhysRevB.95.220406}%
  \BibitemOpen
  \bibfield  {author} {\bibinfo {author} {\bibfnamefont {R.}~\bibnamefont
  {Takagi}}, \bibinfo {author} {\bibfnamefont {D.}~\bibnamefont {Morikawa}},
  \bibinfo {author} {\bibfnamefont {K.}~\bibnamefont {Karube}}, \bibinfo
  {author} {\bibfnamefont {N.}~\bibnamefont {Kanazawa}}, \bibinfo {author}
  {\bibfnamefont {K.}~\bibnamefont {Shibata}}, \bibinfo {author} {\bibfnamefont
  {G.}~\bibnamefont {Tatara}}, \bibinfo {author} {\bibfnamefont
  {Y.}~\bibnamefont {Tokunaga}}, \bibinfo {author} {\bibfnamefont
  {T.}~\bibnamefont {Arima}}, \bibinfo {author} {\bibfnamefont
  {Y.}~\bibnamefont {Taguchi}}, \bibinfo {author} {\bibfnamefont
  {Y.}~\bibnamefont {Tokura}},\ and\ \bibinfo {author} {\bibfnamefont
  {S.}~\bibnamefont {Seki}},\ }\bibfield  {title} {\bibinfo {title} {Spin-wave
  spectroscopy of the $\rm{Dzyaloshinskii}$-$\rm{Moriya}$ interaction in
  room-temperature chiral magnets hosting skyrmions},\ }\href
  {https://doi.org/10.1103/PhysRevB.95.220406} {\bibfield  {journal} {\bibinfo
  {journal} {Phys. Rev. B}\ }\textbf {\bibinfo {volume} {95}},\ \bibinfo
  {pages} {220406(R)} (\bibinfo {year} {2017})}\BibitemShut {NoStop}%
\bibitem [{\citenamefont {Tacchi}\ \emph {et~al.}(2017)\citenamefont {Tacchi},
  \citenamefont {Troncoso}, \citenamefont {Ahlberg}, \citenamefont {Gubbiotti},
  \citenamefont {Madami}, \citenamefont {{\AA}kerman},\ and\ \citenamefont
  {Landeros}}]{tacchi2017interfacial}%
  \BibitemOpen
  \bibfield  {author} {\bibinfo {author} {\bibfnamefont {S.}~\bibnamefont
  {Tacchi}}, \bibinfo {author} {\bibfnamefont {R.~E.}\ \bibnamefont
  {Troncoso}}, \bibinfo {author} {\bibfnamefont {M.}~\bibnamefont {Ahlberg}},
  \bibinfo {author} {\bibfnamefont {G.}~\bibnamefont {Gubbiotti}}, \bibinfo
  {author} {\bibfnamefont {M.}~\bibnamefont {Madami}}, \bibinfo {author}
  {\bibfnamefont {J.}~\bibnamefont {{\AA}kerman}},\ and\ \bibinfo {author}
  {\bibfnamefont {P.}~\bibnamefont {Landeros}},\ }\bibfield  {title} {\bibinfo
  {title} {Interfacial $\rm{Dzyaloshinskii}$-$\rm{Moriya}$ interaction in
  $\rm{Pt/CoFeB}$ films: effect of the heavy-metal thickness},\ }\href@noop {}
  {\bibfield  {journal} {\bibinfo  {journal} {Phys. Rev. Lett.}\ }\textbf
  {\bibinfo {volume} {118}},\ \bibinfo {pages} {147201} (\bibinfo {year}
  {2017})}\BibitemShut {NoStop}%
\bibitem [{\citenamefont {Chaurasiya}\ \emph {et~al.}(2018)\citenamefont
  {Chaurasiya}, \citenamefont {Choudhury}, \citenamefont {Sinha},\ and\
  \citenamefont {Barman}}]{chaurasiya2018dependence}%
  \BibitemOpen
  \bibfield  {author} {\bibinfo {author} {\bibfnamefont {A.~K.}\ \bibnamefont
  {Chaurasiya}}, \bibinfo {author} {\bibfnamefont {S.}~\bibnamefont
  {Choudhury}}, \bibinfo {author} {\bibfnamefont {J.}~\bibnamefont {Sinha}},\
  and\ \bibinfo {author} {\bibfnamefont {A.}~\bibnamefont {Barman}},\
  }\bibfield  {title} {\bibinfo {title} {Dependence of interfacial
  $\rm{Dzyaloshinskii}$-$\rm{Moriya}$ interaction on layer thicknesses in
  $\rm{Ta/Co}$-$\rm{Fe}$-$\rm{B/TaO}$$_x$ heterostructures from brillouin light
  scattering},\ }\href@noop {} {\bibfield  {journal} {\bibinfo  {journal}
  {Phys. Rev. Appl.}\ }\textbf {\bibinfo {volume} {9}},\ \bibinfo {pages}
  {014008} (\bibinfo {year} {2018})}\BibitemShut {NoStop}%
\bibitem [{\citenamefont {Iguchi}\ \emph {et~al.}(2018)\citenamefont {Iguchi},
  \citenamefont {Nii}, \citenamefont {Kawano}, \citenamefont {Murakawa},
  \citenamefont {Hanasaki},\ and\ \citenamefont {Onose}}]{PhysRevB.98.064416}%
  \BibitemOpen
  \bibfield  {author} {\bibinfo {author} {\bibfnamefont {Y.}~\bibnamefont
  {Iguchi}}, \bibinfo {author} {\bibfnamefont {Y.}~\bibnamefont {Nii}},
  \bibinfo {author} {\bibfnamefont {M.}~\bibnamefont {Kawano}}, \bibinfo
  {author} {\bibfnamefont {H.}~\bibnamefont {Murakawa}}, \bibinfo {author}
  {\bibfnamefont {N.}~\bibnamefont {Hanasaki}},\ and\ \bibinfo {author}
  {\bibfnamefont {Y.}~\bibnamefont {Onose}},\ }\bibfield  {title} {\bibinfo
  {title} {Microwave nonreciprocity of magnon excitations in the
  noncentrosymmetric antiferromagnet $\rm{Ba_2MnGe_2O_7}$},\ }\href
  {https://doi.org/10.1103/PhysRevB.98.064416} {\bibfield  {journal} {\bibinfo
  {journal} {Phys. Rev. B}\ }\textbf {\bibinfo {volume} {98}},\ \bibinfo
  {pages} {064416} (\bibinfo {year} {2018})}\BibitemShut {NoStop}%
\bibitem [{\citenamefont {Takahashi}\ \emph {et~al.}(2011)\citenamefont
  {Takahashi}, \citenamefont {Shimano}, \citenamefont {Kaneko}, \citenamefont
  {Murakawa},\ and\ \citenamefont {Tokura}}]{takahashi.Nat.Phys.}%
  \BibitemOpen
  \bibfield  {author} {\bibinfo {author} {\bibfnamefont {Y.}~\bibnamefont
  {Takahashi}}, \bibinfo {author} {\bibfnamefont {R.}~\bibnamefont {Shimano}},
  \bibinfo {author} {\bibfnamefont {Y.}~\bibnamefont {Kaneko}}, \bibinfo
  {author} {\bibfnamefont {H.}~\bibnamefont {Murakawa}},\ and\ \bibinfo
  {author} {\bibfnamefont {Y.}~\bibnamefont {Tokura}},\ }\bibfield  {title}
  {\bibinfo {title} {Magnetoelectric resonance with electromagnons in a
  perovskite helimagnet},\ }\href {https://doi.org/10.1038/nphys2161}
  {\bibfield  {journal} {\bibinfo  {journal} {Nat. Phys.}\ }\textbf {\bibinfo
  {volume} {8}},\ \bibinfo {pages} {121} (\bibinfo {year} {2011})}\BibitemShut
  {NoStop}%
\bibitem [{\citenamefont {Miyahara}\ and\ \citenamefont
  {Furukawa}(2012)}]{doi:10.1143/JPSJ.81.023712}%
  \BibitemOpen
  \bibfield  {author} {\bibinfo {author} {\bibfnamefont {S.}~\bibnamefont
  {Miyahara}}\ and\ \bibinfo {author} {\bibfnamefont {N.}~\bibnamefont
  {Furukawa}},\ }\bibfield  {title} {\bibinfo {title} {Nonreciprocal
  directional dichroism and toroidalmagnons in helical magnets},\ }\href
  {https://doi.org/10.1143/JPSJ.81.023712} {\bibfield  {journal} {\bibinfo
  {journal} {J. Phys. Soc. Jpn.}\ }\textbf {\bibinfo {volume} {81}},\ \bibinfo
  {pages} {023712} (\bibinfo {year} {2012})}\BibitemShut {NoStop}%
\bibitem [{\citenamefont {Miyahara}\ and\ \citenamefont
  {Furukawa}(2013)}]{Miyahara2013}%
  \BibitemOpen
  \bibfield  {author} {\bibinfo {author} {\bibfnamefont {S.}~\bibnamefont
  {Miyahara}}\ and\ \bibinfo {author} {\bibfnamefont {N.}~\bibnamefont
  {Furukawa}},\ }\bibfield  {title} {\bibinfo {title} {Electromagnon in
  multiferroic materials with
  $\rm{Dzyaloshinsky}$-$\rm{Moriya}$-interaction-induced helical spin
  structures},\ }\href {https://doi.org/10.3938/jkps.62.1763} {\bibfield
  {journal} {\bibinfo  {journal} {J. Korean Phys. Soc.}\ }\textbf {\bibinfo
  {volume} {62}},\ \bibinfo {pages} {1763} (\bibinfo {year}
  {2013})}\BibitemShut {NoStop}%
\bibitem [{\citenamefont {Miyahara}\ and\ \citenamefont
  {Furukawa}(2014)}]{PhysRevB.89.195145}%
  \BibitemOpen
  \bibfield  {author} {\bibinfo {author} {\bibfnamefont {S.}~\bibnamefont
  {Miyahara}}\ and\ \bibinfo {author} {\bibfnamefont {N.}~\bibnamefont
  {Furukawa}},\ }\bibfield  {title} {\bibinfo {title} {Theory of
  magneto-optical effects in helical multiferroic materials via toroidal magnon
  excitation},\ }\href {https://doi.org/10.1103/PhysRevB.89.195145} {\bibfield
  {journal} {\bibinfo  {journal} {Phys. Rev. B}\ }\textbf {\bibinfo {volume}
  {89}},\ \bibinfo {pages} {195145} (\bibinfo {year} {2014})}\BibitemShut
  {NoStop}%
\bibitem [{\citenamefont {Mochizuki}(2015)}]{PhysRevLett.114.197203}%
  \BibitemOpen
  \bibfield  {author} {\bibinfo {author} {\bibfnamefont {M.}~\bibnamefont
  {Mochizuki}},\ }\bibfield  {title} {\bibinfo {title} {Microwave magnetochiral
  effect in $\rm{Cu_2OSeO_3}$},\ }\href
  {https://doi.org/10.1103/PhysRevLett.114.197203} {\bibfield  {journal}
  {\bibinfo  {journal} {Phys. Rev. Lett.}\ }\textbf {\bibinfo {volume} {114}},\
  \bibinfo {pages} {197203} (\bibinfo {year} {2015})}\BibitemShut {NoStop}%
\bibitem [{\citenamefont {Proskurin}\ \emph {et~al.}(2018)\citenamefont
  {Proskurin}, \citenamefont {Ovchinnikov}, \citenamefont {Kishine},\ and\
  \citenamefont {Stamps}}]{PhysRevB.98.134422}%
  \BibitemOpen
  \bibfield  {author} {\bibinfo {author} {\bibfnamefont {I.}~\bibnamefont
  {Proskurin}}, \bibinfo {author} {\bibfnamefont {A.~S.}\ \bibnamefont
  {Ovchinnikov}}, \bibinfo {author} {\bibfnamefont {J.-i.}\ \bibnamefont
  {Kishine}},\ and\ \bibinfo {author} {\bibfnamefont {R.~L.}\ \bibnamefont
  {Stamps}},\ }\bibfield  {title} {\bibinfo {title} {Excitation of magnon spin
  photocurrents in antiferromagnetic insulators},\ }\href
  {https://doi.org/10.1103/PhysRevB.98.134422} {\bibfield  {journal} {\bibinfo
  {journal} {Phys. Rev. B}\ }\textbf {\bibinfo {volume} {98}},\ \bibinfo
  {pages} {134422} (\bibinfo {year} {2018})}\BibitemShut {NoStop}%
\bibitem [{\citenamefont {Okuma}(2019)}]{PhysRevB.99.094401}%
  \BibitemOpen
  \bibfield  {author} {\bibinfo {author} {\bibfnamefont {N.}~\bibnamefont
  {Okuma}},\ }\bibfield  {title} {\bibinfo {title} {Nonreciprocal superposition
  state in antiferromagnetic optospintronics},\ }\href
  {https://doi.org/10.1103/PhysRevB.99.094401} {\bibfield  {journal} {\bibinfo
  {journal} {Phys. Rev. B}\ }\textbf {\bibinfo {volume} {99}},\ \bibinfo
  {pages} {094401} (\bibinfo {year} {2019})}\BibitemShut {NoStop}%
\bibitem [{\citenamefont {Takashima}\ \emph {et~al.}(2018)\citenamefont
  {Takashima}, \citenamefont {Shiomi},\ and\ \citenamefont
  {Motome}}]{PhysRevB.98.020401}%
  \BibitemOpen
  \bibfield  {author} {\bibinfo {author} {\bibfnamefont {R.}~\bibnamefont
  {Takashima}}, \bibinfo {author} {\bibfnamefont {Y.}~\bibnamefont {Shiomi}},\
  and\ \bibinfo {author} {\bibfnamefont {Y.}~\bibnamefont {Motome}},\
  }\bibfield  {title} {\bibinfo {title} {Nonreciprocal spin $\rm{Seebeck}$
  effect in antiferromagnets},\ }\href
  {https://doi.org/10.1103/PhysRevB.98.020401} {\bibfield  {journal} {\bibinfo
  {journal} {Phys. Rev. B}\ }\textbf {\bibinfo {volume} {98}},\ \bibinfo
  {pages} {020401(R)} (\bibinfo {year} {2018})}\BibitemShut {NoStop}%
\bibitem [{\citenamefont {Shiomi}\ \emph
  {et~al.}(2017{\natexlab{a}})\citenamefont {Shiomi}, \citenamefont
  {Takashima}, \citenamefont {Okuyama}, \citenamefont {Gitgeatpong},
  \citenamefont {Piyawongwatthana}, \citenamefont {Matan}, \citenamefont
  {Sato},\ and\ \citenamefont {Saitoh}}]{PhysRevB.96.180414}%
  \BibitemOpen
  \bibfield  {author} {\bibinfo {author} {\bibfnamefont {Y.}~\bibnamefont
  {Shiomi}}, \bibinfo {author} {\bibfnamefont {R.}~\bibnamefont {Takashima}},
  \bibinfo {author} {\bibfnamefont {D.}~\bibnamefont {Okuyama}}, \bibinfo
  {author} {\bibfnamefont {G.}~\bibnamefont {Gitgeatpong}}, \bibinfo {author}
  {\bibfnamefont {P.}~\bibnamefont {Piyawongwatthana}}, \bibinfo {author}
  {\bibfnamefont {K.}~\bibnamefont {Matan}}, \bibinfo {author} {\bibfnamefont
  {T.~J.}\ \bibnamefont {Sato}},\ and\ \bibinfo {author} {\bibfnamefont
  {E.}~\bibnamefont {Saitoh}},\ }\bibfield  {title} {\bibinfo {title} {Spin
  $\rm{Seebeck}$ effect in the polar antiferromagnet
  $\alpha$-$\rm{Cu_2V_2O_7}$},\ }\href
  {https://doi.org/10.1103/PhysRevB.96.180414} {\bibfield  {journal} {\bibinfo
  {journal} {Phys. Rev. B}\ }\textbf {\bibinfo {volume} {96}},\ \bibinfo
  {pages} {180414(R)} (\bibinfo {year} {2017}{\natexlab{a}})}\BibitemShut
  {NoStop}%
\bibitem [{\citenamefont {Gr{\"u}nberg}\ \emph {et~al.}(1986)\citenamefont
  {Gr{\"u}nberg}, \citenamefont {Schreiber}, \citenamefont {Pang},
  \citenamefont {Brodsky},\ and\ \citenamefont {Sowers}}]{grunberg1986layered}%
  \BibitemOpen
  \bibfield  {author} {\bibinfo {author} {\bibfnamefont {P.}~\bibnamefont
  {Gr{\"u}nberg}}, \bibinfo {author} {\bibfnamefont {R.}~\bibnamefont
  {Schreiber}}, \bibinfo {author} {\bibfnamefont {Y.}~\bibnamefont {Pang}},
  \bibinfo {author} {\bibfnamefont {M.~B.}\ \bibnamefont {Brodsky}},\ and\
  \bibinfo {author} {\bibfnamefont {H.}~\bibnamefont {Sowers}},\ }\bibfield
  {title} {\bibinfo {title} {Layered magnetic structures: Evidence for
  antiferromagnetic coupling of $\rm{Fe}$ layers across $\rm{Cr}$
  interlayers},\ }\href@noop {} {\bibfield  {journal} {\bibinfo  {journal}
  {Phys. Rev. Lett.}\ }\textbf {\bibinfo {volume} {57}},\ \bibinfo {pages}
  {2442} (\bibinfo {year} {1986})}\BibitemShut {NoStop}%
\bibitem [{\citenamefont {Zhang}\ and\ \citenamefont
  {Zinn}(1987)}]{zhang1987spin}%
  \BibitemOpen
  \bibfield  {author} {\bibinfo {author} {\bibfnamefont {P.~X.}\ \bibnamefont
  {Zhang}}\ and\ \bibinfo {author} {\bibfnamefont {W.}~\bibnamefont {Zinn}},\
  }\bibfield  {title} {\bibinfo {title} {Spin-wave modes in antiparallel
  magnetized ferromagnetic double layers},\ }\href@noop {} {\bibfield
  {journal} {\bibinfo  {journal} {Physical Review B}\ }\textbf {\bibinfo
  {volume} {35}},\ \bibinfo {pages} {5219} (\bibinfo {year}
  {1987})}\BibitemShut {NoStop}%
\bibitem [{\citenamefont {Di}\ \emph {et~al.}(2015)\citenamefont {Di},
  \citenamefont {Feng}, \citenamefont {Piramanayagam}, \citenamefont {Zhang},
  \citenamefont {Lim}, \citenamefont {Ng},\ and\ \citenamefont
  {Kuok}}]{di2015enhancement}%
  \BibitemOpen
  \bibfield  {author} {\bibinfo {author} {\bibfnamefont {K.}~\bibnamefont
  {Di}}, \bibinfo {author} {\bibfnamefont {S.}~\bibnamefont {Feng}}, \bibinfo
  {author} {\bibfnamefont {S.~N.}\ \bibnamefont {Piramanayagam}}, \bibinfo
  {author} {\bibfnamefont {V.}~\bibnamefont {Zhang}}, \bibinfo {author}
  {\bibfnamefont {H.~S.}\ \bibnamefont {Lim}}, \bibinfo {author} {\bibfnamefont
  {S.~C.}\ \bibnamefont {Ng}},\ and\ \bibinfo {author} {\bibfnamefont {M.~H.}\
  \bibnamefont {Kuok}},\ }\bibfield  {title} {\bibinfo {title} {Enhancement of
  spin-wave nonreciprocity in magnonic crystals via synthetic antiferromagnetic
  coupling},\ }\href@noop {} {\bibfield  {journal} {\bibinfo  {journal} {Sci.
  Rep.}\ }\textbf {\bibinfo {volume} {5}},\ \bibinfo {pages} {10153} (\bibinfo
  {year} {2015})}\BibitemShut {NoStop}%
\bibitem [{\citenamefont {Gallardo}\ \emph
  {et~al.}(2019{\natexlab{a}})\citenamefont {Gallardo}, \citenamefont
  {Schneider}, \citenamefont {Chaurasiya}, \citenamefont {Oelschl\"agel},
  \citenamefont {Arekapudi}, \citenamefont {Rold\'an-Molina}, \citenamefont
  {H\"ubner}, \citenamefont {Lenz}, \citenamefont {Barman}, \citenamefont
  {Fassbender}, \citenamefont {Lindner}, \citenamefont {Hellwig},\ and\
  \citenamefont {Landeros}}]{gallardo2019reconfigurable}%
  \BibitemOpen
  \bibfield  {author} {\bibinfo {author} {\bibfnamefont {R.~A.}\ \bibnamefont
  {Gallardo}}, \bibinfo {author} {\bibfnamefont {T.}~\bibnamefont {Schneider}},
  \bibinfo {author} {\bibfnamefont {A.~K.}\ \bibnamefont {Chaurasiya}},
  \bibinfo {author} {\bibfnamefont {A.}~\bibnamefont {Oelschl\"agel}}, \bibinfo
  {author} {\bibfnamefont {S.~S. P.~K.}\ \bibnamefont {Arekapudi}}, \bibinfo
  {author} {\bibfnamefont {A.}~\bibnamefont {Rold\'an-Molina}}, \bibinfo
  {author} {\bibfnamefont {R.}~\bibnamefont {H\"ubner}}, \bibinfo {author}
  {\bibfnamefont {K.}~\bibnamefont {Lenz}}, \bibinfo {author} {\bibfnamefont
  {A.}~\bibnamefont {Barman}}, \bibinfo {author} {\bibfnamefont
  {J.}~\bibnamefont {Fassbender}}, \bibinfo {author} {\bibfnamefont
  {J.}~\bibnamefont {Lindner}}, \bibinfo {author} {\bibfnamefont
  {O.}~\bibnamefont {Hellwig}},\ and\ \bibinfo {author} {\bibfnamefont
  {P.}~\bibnamefont {Landeros}},\ }\bibfield  {title} {\bibinfo {title}
  {Reconfigurable spin-wave nonreciprocity induced by dipolar interaction in a
  coupled ferromagnetic bilayer},\ }\href@noop {} {\bibfield  {journal}
  {\bibinfo  {journal} {Phys. Rev. Appl.}\ }\textbf {\bibinfo {volume} {12}},\
  \bibinfo {pages} {034012} (\bibinfo {year} {2019}{\natexlab{a}})}\BibitemShut
  {NoStop}%
\bibitem [{\citenamefont {Albisetti}\ \emph {et~al.}(2020)\citenamefont
  {Albisetti}, \citenamefont {Tacchi}, \citenamefont {Silvani}, \citenamefont
  {Scaramuzzi}, \citenamefont {Finizio}, \citenamefont {Wintz}, \citenamefont
  {Rinaldi}, \citenamefont {Cantoni}, \citenamefont {Raabe}, \citenamefont
  {Carlotti} \emph {et~al.}}]{albisetti2020optically}%
  \BibitemOpen
  \bibfield  {author} {\bibinfo {author} {\bibfnamefont {E.}~\bibnamefont
  {Albisetti}}, \bibinfo {author} {\bibfnamefont {S.}~\bibnamefont {Tacchi}},
  \bibinfo {author} {\bibfnamefont {R.}~\bibnamefont {Silvani}}, \bibinfo
  {author} {\bibfnamefont {G.}~\bibnamefont {Scaramuzzi}}, \bibinfo {author}
  {\bibfnamefont {S.}~\bibnamefont {Finizio}}, \bibinfo {author} {\bibfnamefont
  {S.}~\bibnamefont {Wintz}}, \bibinfo {author} {\bibfnamefont
  {C.}~\bibnamefont {Rinaldi}}, \bibinfo {author} {\bibfnamefont
  {M.}~\bibnamefont {Cantoni}}, \bibinfo {author} {\bibfnamefont
  {J.}~\bibnamefont {Raabe}}, \bibinfo {author} {\bibfnamefont
  {G.}~\bibnamefont {Carlotti}}, \emph {et~al.},\ }\bibfield  {title} {\bibinfo
  {title} {Optically inspired nanomagnonics with nonreciprocal spin waves in
  synthetic antiferromagnets},\ }\href@noop {} {\bibfield  {journal} {\bibinfo
  {journal} {Adv. Mater.}\ ,\ \bibinfo {pages} {1906439}} (\bibinfo {year}
  {2020})}\BibitemShut {NoStop}%
\bibitem [{\citenamefont {Cheon}\ \emph {et~al.}(2018)\citenamefont {Cheon},
  \citenamefont {Lee},\ and\ \citenamefont {Cheong}}]{PhysRevB.98.184405}%
  \BibitemOpen
  \bibfield  {author} {\bibinfo {author} {\bibfnamefont {S.}~\bibnamefont
  {Cheon}}, \bibinfo {author} {\bibfnamefont {H.-W.}\ \bibnamefont {Lee}},\
  and\ \bibinfo {author} {\bibfnamefont {S.-W.}\ \bibnamefont {Cheong}},\
  }\bibfield  {title} {\bibinfo {title} {Nonreciprocal spin waves in a chiral
  antiferromagnet without the $\rm{Dzyaloshinskii}$-$\rm{Moriya}$
  interaction},\ }\href {https://doi.org/10.1103/PhysRevB.98.184405} {\bibfield
   {journal} {\bibinfo  {journal} {Phys. Rev. B}\ }\textbf {\bibinfo {volume}
  {98}},\ \bibinfo {pages} {184405} (\bibinfo {year} {2018})}\BibitemShut
  {NoStop}%
\bibitem [{\citenamefont {Maksimov}\ \emph {et~al.}(2019)\citenamefont
  {Maksimov}, \citenamefont {Zhu}, \citenamefont {White},\ and\ \citenamefont
  {Chernyshev}}]{maksimov2019anisotropic}%
  \BibitemOpen
  \bibfield  {author} {\bibinfo {author} {\bibfnamefont {P.~A.}\ \bibnamefont
  {Maksimov}}, \bibinfo {author} {\bibfnamefont {Z.}~\bibnamefont {Zhu}},
  \bibinfo {author} {\bibfnamefont {S.~R.}\ \bibnamefont {White}},\ and\
  \bibinfo {author} {\bibfnamefont {A.~L.}\ \bibnamefont {Chernyshev}},\
  }\bibfield  {title} {\bibinfo {title} {Anisotropic-exchange magnets on a
  triangular lattice: spin waves, accidental degeneracies, and dual spin
  liquids},\ }\href@noop {} {\bibfield  {journal} {\bibinfo  {journal}
  {Physical Review X}\ }\textbf {\bibinfo {volume} {9}},\ \bibinfo {pages}
  {021017} (\bibinfo {year} {2019})}\BibitemShut {NoStop}%
\bibitem [{\citenamefont {Ot{\'a}lora}\ \emph {et~al.}(2016)\citenamefont
  {Ot{\'a}lora}, \citenamefont {Yan}, \citenamefont {Schultheiss},
  \citenamefont {Hertel},\ and\ \citenamefont
  {K{\'a}kay}}]{otalora2016curvature}%
  \BibitemOpen
  \bibfield  {author} {\bibinfo {author} {\bibfnamefont {J.~A.}\ \bibnamefont
  {Ot{\'a}lora}}, \bibinfo {author} {\bibfnamefont {M.}~\bibnamefont {Yan}},
  \bibinfo {author} {\bibfnamefont {H.}~\bibnamefont {Schultheiss}}, \bibinfo
  {author} {\bibfnamefont {R.}~\bibnamefont {Hertel}},\ and\ \bibinfo {author}
  {\bibfnamefont {A.}~\bibnamefont {K{\'a}kay}},\ }\bibfield  {title} {\bibinfo
  {title} {Curvature-induced asymmetric spin-wave dispersion},\ }\href@noop {}
  {\bibfield  {journal} {\bibinfo  {journal} {Phys. Rev. Lett.}\ }\textbf
  {\bibinfo {volume} {117}},\ \bibinfo {pages} {227203} (\bibinfo {year}
  {2016})}\BibitemShut {NoStop}%
\bibitem [{\citenamefont {Gallardo}\ \emph
  {et~al.}(2019{\natexlab{b}})\citenamefont {Gallardo}, \citenamefont
  {Alvarado-Seguel}, \citenamefont {Schneider}, \citenamefont
  {Gonzalez-Fuentes}, \citenamefont {Rold{\'a}n-Molina}, \citenamefont {Lenz},
  \citenamefont {Lindner},\ and\ \citenamefont {Landeros}}]{gallardo2019spin}%
  \BibitemOpen
  \bibfield  {author} {\bibinfo {author} {\bibfnamefont {R.}~\bibnamefont
  {Gallardo}}, \bibinfo {author} {\bibfnamefont {P.}~\bibnamefont
  {Alvarado-Seguel}}, \bibinfo {author} {\bibfnamefont {T.}~\bibnamefont
  {Schneider}}, \bibinfo {author} {\bibfnamefont {C.}~\bibnamefont
  {Gonzalez-Fuentes}}, \bibinfo {author} {\bibfnamefont {A.}~\bibnamefont
  {Rold{\'a}n-Molina}}, \bibinfo {author} {\bibfnamefont {K.}~\bibnamefont
  {Lenz}}, \bibinfo {author} {\bibfnamefont {J.}~\bibnamefont {Lindner}},\ and\
  \bibinfo {author} {\bibfnamefont {P.}~\bibnamefont {Landeros}},\ }\bibfield
  {title} {\bibinfo {title} {Spin-wave non-reciprocity in magnetization-graded
  ferromagnetic films},\ }\href@noop {} {\bibfield  {journal} {\bibinfo
  {journal} {New J. Phys.}\ }\textbf {\bibinfo {volume} {21}},\ \bibinfo
  {pages} {033026} (\bibinfo {year} {2019}{\natexlab{b}})}\BibitemShut
  {NoStop}%
\bibitem [{\citenamefont {Li}\ \emph {et~al.}(2016)\citenamefont {Li},
  \citenamefont {Wang},\ and\ \citenamefont {Chen}}]{PhysRevB.94.035107}%
  \BibitemOpen
  \bibfield  {author} {\bibinfo {author} {\bibfnamefont {Y.-D.}\ \bibnamefont
  {Li}}, \bibinfo {author} {\bibfnamefont {X.}~\bibnamefont {Wang}},\ and\
  \bibinfo {author} {\bibfnamefont {G.}~\bibnamefont {Chen}},\ }\bibfield
  {title} {\bibinfo {title} {Anisotropic spin model of strong
  spin-orbit-coupled triangular antiferromagnets},\ }\href
  {https://doi.org/10.1103/PhysRevB.94.035107} {\bibfield  {journal} {\bibinfo
  {journal} {Phys. Rev. B}\ }\textbf {\bibinfo {volume} {94}},\ \bibinfo
  {pages} {035107} (\bibinfo {year} {2016})}\BibitemShut {NoStop}%
\bibitem [{\citenamefont {Wang}\ \emph {et~al.}(2017)\citenamefont {Wang},
  \citenamefont {Su},\ and\ \citenamefont {Wang}}]{PhysRevB.95.014435}%
  \BibitemOpen
  \bibfield  {author} {\bibinfo {author} {\bibfnamefont {X.~S.}\ \bibnamefont
  {Wang}}, \bibinfo {author} {\bibfnamefont {Y.}~\bibnamefont {Su}},\ and\
  \bibinfo {author} {\bibfnamefont {X.~R.}\ \bibnamefont {Wang}},\ }\bibfield
  {title} {\bibinfo {title} {Topologically protected unidirectional edge spin
  waves and beam splitter},\ }\href
  {https://doi.org/10.1103/PhysRevB.95.014435} {\bibfield  {journal} {\bibinfo
  {journal} {Phys. Rev. B}\ }\textbf {\bibinfo {volume} {95}},\ \bibinfo
  {pages} {014435} (\bibinfo {year} {2017})}\BibitemShut {NoStop}%
\bibitem [{\citenamefont {Wang}\ \emph {et~al.}(2018)\citenamefont {Wang},
  \citenamefont {Zhang},\ and\ \citenamefont {Wang}}]{PhysRevApplied.9.024029}%
  \BibitemOpen
  \bibfield  {author} {\bibinfo {author} {\bibfnamefont {X.~S.}\ \bibnamefont
  {Wang}}, \bibinfo {author} {\bibfnamefont {H.~W.}\ \bibnamefont {Zhang}},\
  and\ \bibinfo {author} {\bibfnamefont {X.~R.}\ \bibnamefont {Wang}},\
  }\bibfield  {title} {\bibinfo {title} {Topological magnonics: A paradigm for
  spin-wave manipulation and device design},\ }\href
  {https://doi.org/10.1103/PhysRevApplied.9.024029} {\bibfield  {journal}
  {\bibinfo  {journal} {Phys. Rev. Applied}\ }\textbf {\bibinfo {volume} {9}},\
  \bibinfo {pages} {024029} (\bibinfo {year} {2018})}\BibitemShut {NoStop}%
\bibitem [{\citenamefont {Colpa}(1978)}]{COLPA1978327}%
  \BibitemOpen
  \bibfield  {author} {\bibinfo {author} {\bibfnamefont {J.}~\bibnamefont
  {Colpa}},\ }\bibfield  {title} {\bibinfo {title} {Diagonalization of the
  quadratic boson hamiltonian},\ }\href
  {https://doi.org/https://doi.org/10.1016/0378-4371(78)90160-7} {\bibfield
  {journal} {\bibinfo  {journal} {Physica A}\ }\textbf {\bibinfo {volume}
  {93}},\ \bibinfo {pages} {327 } (\bibinfo {year} {1978})}\BibitemShut
  {NoStop}%
\bibitem [{\citenamefont {Spaldin}\ \emph {et~al.}(2008)\citenamefont
  {Spaldin}, \citenamefont {Fiebig},\ and\ \citenamefont
  {Mostovoy}}]{Spaldin_0953-8984-20-43-434203}%
  \BibitemOpen
  \bibfield  {author} {\bibinfo {author} {\bibfnamefont {N.~A.}\ \bibnamefont
  {Spaldin}}, \bibinfo {author} {\bibfnamefont {M.}~\bibnamefont {Fiebig}},\
  and\ \bibinfo {author} {\bibfnamefont {M.}~\bibnamefont {Mostovoy}},\
  }\bibfield  {title} {\bibinfo {title} {The toroidal moment in
  condensed-matter physics and its relation to the magnetoelectric effect},\
  }\href@noop {} {\bibfield  {journal} {\bibinfo  {journal} {J. Phys.: Condens.
  Matter}\ }\textbf {\bibinfo {volume} {20}},\ \bibinfo {pages} {434203}
  (\bibinfo {year} {2008})}\BibitemShut {NoStop}%
\bibitem [{\citenamefont {Hayami}\ and\ \citenamefont
  {Kusunose}(2018)}]{hayami2018microscopic}%
  \BibitemOpen
  \bibfield  {author} {\bibinfo {author} {\bibfnamefont {S.}~\bibnamefont
  {Hayami}}\ and\ \bibinfo {author} {\bibfnamefont {H.}~\bibnamefont
  {Kusunose}},\ }\bibfield  {title} {\bibinfo {title} {Microscopic description
  of electric and magnetic toroidal multipoles in hybrid orbitals},\
  }\href@noop {} {\bibfield  {journal} {\bibinfo  {journal} {Journal of the
  Physical Society of Japan}\ }\textbf {\bibinfo {volume} {87}},\ \bibinfo
  {pages} {033709} (\bibinfo {year} {2018})}\BibitemShut {NoStop}%
\bibitem [{\citenamefont {Hayami}\ \emph {et~al.}(2018)\citenamefont {Hayami},
  \citenamefont {Yatsushiro}, \citenamefont {Yanagi},\ and\ \citenamefont
  {Kusunose}}]{PhysRevB.98.165110}%
  \BibitemOpen
  \bibfield  {author} {\bibinfo {author} {\bibfnamefont {S.}~\bibnamefont
  {Hayami}}, \bibinfo {author} {\bibfnamefont {M.}~\bibnamefont {Yatsushiro}},
  \bibinfo {author} {\bibfnamefont {Y.}~\bibnamefont {Yanagi}},\ and\ \bibinfo
  {author} {\bibfnamefont {H.}~\bibnamefont {Kusunose}},\ }\bibfield  {title}
  {\bibinfo {title} {Classification of atomic-scale multipoles under
  crystallographic point groups and application to linear response tensors},\
  }\href {https://doi.org/10.1103/PhysRevB.98.165110} {\bibfield  {journal}
  {\bibinfo  {journal} {Phys. Rev. B}\ }\textbf {\bibinfo {volume} {98}},\
  \bibinfo {pages} {165110} (\bibinfo {year} {2018})}\BibitemShut {NoStop}%
\bibitem [{\citenamefont {Suzuki}\ \emph {et~al.}(2019)\citenamefont {Suzuki},
  \citenamefont {Nomoto}, \citenamefont {Arita}, \citenamefont {Yanagi},
  \citenamefont {Hayami},\ and\ \citenamefont {Kusunose}}]{PhysRevB.99.174407}%
  \BibitemOpen
  \bibfield  {author} {\bibinfo {author} {\bibfnamefont {M.-T.}\ \bibnamefont
  {Suzuki}}, \bibinfo {author} {\bibfnamefont {T.}~\bibnamefont {Nomoto}},
  \bibinfo {author} {\bibfnamefont {R.}~\bibnamefont {Arita}}, \bibinfo
  {author} {\bibfnamefont {Y.}~\bibnamefont {Yanagi}}, \bibinfo {author}
  {\bibfnamefont {S.}~\bibnamefont {Hayami}},\ and\ \bibinfo {author}
  {\bibfnamefont {H.}~\bibnamefont {Kusunose}},\ }\bibfield  {title} {\bibinfo
  {title} {Multipole expansion for magnetic structures: A generation scheme for
  a symmetry-adapted orthonormal basis set in the crystallographic point
  group},\ }\href {https://doi.org/10.1103/PhysRevB.99.174407} {\bibfield
  {journal} {\bibinfo  {journal} {Phys. Rev. B}\ }\textbf {\bibinfo {volume}
  {99}},\ \bibinfo {pages} {174407} (\bibinfo {year} {2019})}\BibitemShut
  {NoStop}%
\bibitem [{\citenamefont {Matsumoto}\ \emph {et~al.}(2017)\citenamefont
  {Matsumoto}, \citenamefont {Chimata},\ and\ \citenamefont
  {Koga}}]{matsumoto2017symmetry}%
  \BibitemOpen
  \bibfield  {author} {\bibinfo {author} {\bibfnamefont {M.}~\bibnamefont
  {Matsumoto}}, \bibinfo {author} {\bibfnamefont {K.}~\bibnamefont {Chimata}},\
  and\ \bibinfo {author} {\bibfnamefont {M.}~\bibnamefont {Koga}},\ }\bibfield
  {title} {\bibinfo {title} {Symmetry analysis of spin-dependent electric
  dipole and its application to magnetoelectric effects},\ }\href@noop {}
  {\bibfield  {journal} {\bibinfo  {journal} {Journal of the Physical Society
  of Japan}\ }\textbf {\bibinfo {volume} {86}},\ \bibinfo {pages} {034704}
  (\bibinfo {year} {2017})}\BibitemShut {NoStop}%
\bibitem [{\citenamefont {Khanh}\ \emph {et~al.}(2016)\citenamefont {Khanh},
  \citenamefont {Abe}, \citenamefont {Sagayama}, \citenamefont {Nakao},
  \citenamefont {Hanashima}, \citenamefont {Kiyanagi}, \citenamefont
  {Tokunaga},\ and\ \citenamefont {Arima}}]{Khanh_PhysRevB.93.075117}%
  \BibitemOpen
  \bibfield  {author} {\bibinfo {author} {\bibfnamefont {N.~D.}\ \bibnamefont
  {Khanh}}, \bibinfo {author} {\bibfnamefont {N.}~\bibnamefont {Abe}}, \bibinfo
  {author} {\bibfnamefont {H.}~\bibnamefont {Sagayama}}, \bibinfo {author}
  {\bibfnamefont {A.}~\bibnamefont {Nakao}}, \bibinfo {author} {\bibfnamefont
  {T.}~\bibnamefont {Hanashima}}, \bibinfo {author} {\bibfnamefont
  {R.}~\bibnamefont {Kiyanagi}}, \bibinfo {author} {\bibfnamefont
  {Y.}~\bibnamefont {Tokunaga}},\ and\ \bibinfo {author} {\bibfnamefont
  {T.}~\bibnamefont {Arima}},\ }\bibfield  {title} {\bibinfo {title}
  {Magnetoelectric coupling in the honeycomb antiferromagnet
  $\rm{Co_4Nb_2O_9}$},\ }\href {https://doi.org/10.1103/PhysRevB.93.075117}
  {\bibfield  {journal} {\bibinfo  {journal} {Phys. Rev. B}\ }\textbf {\bibinfo
  {volume} {93}},\ \bibinfo {pages} {075117} (\bibinfo {year}
  {2016})}\BibitemShut {NoStop}%
\bibitem [{\citenamefont {Khanh}\ \emph {et~al.}(2017)\citenamefont {Khanh},
  \citenamefont {Abe}, \citenamefont {Kimura}, \citenamefont {Tokunaga},\ and\
  \citenamefont {Arima}}]{Khanh_PhysRevB.96.094434}%
  \BibitemOpen
  \bibfield  {author} {\bibinfo {author} {\bibfnamefont {N.~D.}\ \bibnamefont
  {Khanh}}, \bibinfo {author} {\bibfnamefont {N.}~\bibnamefont {Abe}}, \bibinfo
  {author} {\bibfnamefont {S.}~\bibnamefont {Kimura}}, \bibinfo {author}
  {\bibfnamefont {Y.}~\bibnamefont {Tokunaga}},\ and\ \bibinfo {author}
  {\bibfnamefont {T.}~\bibnamefont {Arima}},\ }\bibfield  {title} {\bibinfo
  {title} {Manipulation of electric polarization with rotating magnetic field
  in a honeycomb antiferromagnet $\rm{Co_4Nb_2O_9}$},\ }\href
  {https://doi.org/10.1103/PhysRevB.96.094434} {\bibfield  {journal} {\bibinfo
  {journal} {Phys. Rev. B}\ }\textbf {\bibinfo {volume} {96}},\ \bibinfo
  {pages} {094434} (\bibinfo {year} {2017})}\BibitemShut {NoStop}%
\bibitem [{\citenamefont {Yanagi}\ \emph {et~al.}(2018)\citenamefont {Yanagi},
  \citenamefont {Hayami},\ and\ \citenamefont
  {Kusunose}}]{Yanagi_PhysRevB.97.020404}%
  \BibitemOpen
  \bibfield  {author} {\bibinfo {author} {\bibfnamefont {Y.}~\bibnamefont
  {Yanagi}}, \bibinfo {author} {\bibfnamefont {S.}~\bibnamefont {Hayami}},\
  and\ \bibinfo {author} {\bibfnamefont {H.}~\bibnamefont {Kusunose}},\
  }\bibfield  {title} {\bibinfo {title} {Manipulating the magnetoelectric
  effect: Essence learned from $\rm{Co_4Nb_2O_9}$},\ }\href
  {https://doi.org/10.1103/PhysRevB.97.020404} {\bibfield  {journal} {\bibinfo
  {journal} {Phys. Rev. B}\ }\textbf {\bibinfo {volume} {97}},\ \bibinfo
  {pages} {020404(R)} (\bibinfo {year} {2018})}\BibitemShut {NoStop}%
\bibitem [{\citenamefont {Matsumoto}\ and\ \citenamefont
  {Koga}(2019)}]{doi:10.7566/JPSJ.88.094704}%
  \BibitemOpen
  \bibfield  {author} {\bibinfo {author} {\bibfnamefont {M.}~\bibnamefont
  {Matsumoto}}\ and\ \bibinfo {author} {\bibfnamefont {M.}~\bibnamefont
  {Koga}},\ }\bibfield  {title} {\bibinfo {title} {Symmetry analysis of
  magnetoelectric effects in honeycomb antiferromagnet $\rm{Co_4Nb_2O_9}$},\
  }\href {https://doi.org/10.7566/JPSJ.88.094704} {\bibfield  {journal}
  {\bibinfo  {journal} {J. Phys. Soc. Jpn.}\ }\textbf {\bibinfo {volume}
  {88}},\ \bibinfo {pages} {094704} (\bibinfo {year} {2019})}\BibitemShut
  {NoStop}%
\bibitem [{\citenamefont {Jackeli}\ and\ \citenamefont
  {Khaliullin}(2009)}]{PhysRevLett.102.017205}%
  \BibitemOpen
  \bibfield  {author} {\bibinfo {author} {\bibfnamefont {G.}~\bibnamefont
  {Jackeli}}\ and\ \bibinfo {author} {\bibfnamefont {G.}~\bibnamefont
  {Khaliullin}},\ }\bibfield  {title} {\bibinfo {title} {Mott insulators in the
  strong spin-orbit coupling limit: From $\rm{Heisenberg}$ to a quantum compass
  and $\rm{Kitaev}$ models},\ }\href
  {https://doi.org/10.1103/PhysRevLett.102.017205} {\bibfield  {journal}
  {\bibinfo  {journal} {Phys. Rev. Lett.}\ }\textbf {\bibinfo {volume} {102}},\
  \bibinfo {pages} {017205} (\bibinfo {year} {2009})}\BibitemShut {NoStop}%
\bibitem [{\citenamefont {Ressouche}\ \emph {et~al.}(2010)\citenamefont
  {Ressouche}, \citenamefont {Loire}, \citenamefont {Simonet}, \citenamefont
  {Ballou}, \citenamefont {Stunault},\ and\ \citenamefont
  {Wildes}}]{PhysRevB.82.100408}%
  \BibitemOpen
  \bibfield  {author} {\bibinfo {author} {\bibfnamefont {E.}~\bibnamefont
  {Ressouche}}, \bibinfo {author} {\bibfnamefont {M.}~\bibnamefont {Loire}},
  \bibinfo {author} {\bibfnamefont {V.}~\bibnamefont {Simonet}}, \bibinfo
  {author} {\bibfnamefont {R.}~\bibnamefont {Ballou}}, \bibinfo {author}
  {\bibfnamefont {A.}~\bibnamefont {Stunault}},\ and\ \bibinfo {author}
  {\bibfnamefont {A.}~\bibnamefont {Wildes}},\ }\bibfield  {title} {\bibinfo
  {title} {Magnetoelectric $\rm{MnPS_3}$ as a candidate for ferrotoroidicity},\
  }\href {https://doi.org/10.1103/PhysRevB.82.100408} {\bibfield  {journal}
  {\bibinfo  {journal} {Phys. Rev. B}\ }\textbf {\bibinfo {volume} {82}},\
  \bibinfo {pages} {100408(R)} (\bibinfo {year} {2010})}\BibitemShut {NoStop}%
\bibitem [{\citenamefont {Li}\ \emph {et~al.}(2013)\citenamefont {Li},
  \citenamefont {Cao}, \citenamefont {Niu}, \citenamefont {Shi},\ and\
  \citenamefont {Feng}}]{Li3738}%
  \BibitemOpen
  \bibfield  {author} {\bibinfo {author} {\bibfnamefont {X.}~\bibnamefont
  {Li}}, \bibinfo {author} {\bibfnamefont {T.}~\bibnamefont {Cao}}, \bibinfo
  {author} {\bibfnamefont {Q.}~\bibnamefont {Niu}}, \bibinfo {author}
  {\bibfnamefont {J.}~\bibnamefont {Shi}},\ and\ \bibinfo {author}
  {\bibfnamefont {J.}~\bibnamefont {Feng}},\ }\bibfield  {title} {\bibinfo
  {title} {Coupling the valley degree of freedom to antiferromagnetic order},\
  }\href {https://doi.org/10.1073/pnas.1219420110} {\bibfield  {journal}
  {\bibinfo  {journal} {Proc. Natl. Acad. Sci.}\ }\textbf {\bibinfo {volume}
  {110}},\ \bibinfo {pages} {3738} (\bibinfo {year} {2013})}\BibitemShut
  {NoStop}%
\bibitem [{\citenamefont {Sivadas}\ \emph {et~al.}(2015)\citenamefont
  {Sivadas}, \citenamefont {Daniels}, \citenamefont {Swendsen}, \citenamefont
  {Okamoto},\ and\ \citenamefont {Xiao}}]{PhysRevB.91.235425}%
  \BibitemOpen
  \bibfield  {author} {\bibinfo {author} {\bibfnamefont {N.}~\bibnamefont
  {Sivadas}}, \bibinfo {author} {\bibfnamefont {M.~W.}\ \bibnamefont
  {Daniels}}, \bibinfo {author} {\bibfnamefont {R.~H.}\ \bibnamefont
  {Swendsen}}, \bibinfo {author} {\bibfnamefont {S.}~\bibnamefont {Okamoto}},\
  and\ \bibinfo {author} {\bibfnamefont {D.}~\bibnamefont {Xiao}},\ }\bibfield
  {title} {\bibinfo {title} {Magnetic ground state of semiconducting
  transition-metal trichalcogenide monolayers},\ }\href
  {https://doi.org/10.1103/PhysRevB.91.235425} {\bibfield  {journal} {\bibinfo
  {journal} {Phys. Rev. B}\ }\textbf {\bibinfo {volume} {91}},\ \bibinfo
  {pages} {235425} (\bibinfo {year} {2015})}\BibitemShut {NoStop}%
\bibitem [{\citenamefont {Shiomi}\ \emph
  {et~al.}(2017{\natexlab{b}})\citenamefont {Shiomi}, \citenamefont
  {Takashima},\ and\ \citenamefont {Saitoh}}]{PhysRevB.96.134425}%
  \BibitemOpen
  \bibfield  {author} {\bibinfo {author} {\bibfnamefont {Y.}~\bibnamefont
  {Shiomi}}, \bibinfo {author} {\bibfnamefont {R.}~\bibnamefont {Takashima}},\
  and\ \bibinfo {author} {\bibfnamefont {E.}~\bibnamefont {Saitoh}},\
  }\bibfield  {title} {\bibinfo {title} {Experimental evidence consistent with
  a magnon $\rm{Nernst}$ effect in the antiferromagnetic insulator
  $\rm{MnPS_3}$},\ }\href {https://doi.org/10.1103/PhysRevB.96.134425}
  {\bibfield  {journal} {\bibinfo  {journal} {Phys. Rev. B}\ }\textbf {\bibinfo
  {volume} {96}},\ \bibinfo {pages} {134425} (\bibinfo {year}
  {2017}{\natexlab{b}})}\BibitemShut {NoStop}%
\bibitem [{\citenamefont {Ninomiya}\ \emph {et~al.}(2018)\citenamefont
  {Ninomiya}, \citenamefont {Sato}, \citenamefont {Matsumoto}, \citenamefont
  {Moyoshi}, \citenamefont {Nakao}, \citenamefont {Ohishi}, \citenamefont
  {Kousaka}, \citenamefont {Akimitsu}, \citenamefont {Inoue},\ and\
  \citenamefont {Ohara}}]{e1ddc90bb22e4448a040d6ce3fdca500}%
  \BibitemOpen
  \bibfield  {author} {\bibinfo {author} {\bibfnamefont {H.}~\bibnamefont
  {Ninomiya}}, \bibinfo {author} {\bibfnamefont {T.}~\bibnamefont {Sato}},
  \bibinfo {author} {\bibfnamefont {Y.}~\bibnamefont {Matsumoto}}, \bibinfo
  {author} {\bibfnamefont {T.}~\bibnamefont {Moyoshi}}, \bibinfo {author}
  {\bibfnamefont {A.}~\bibnamefont {Nakao}}, \bibinfo {author} {\bibfnamefont
  {K.}~\bibnamefont {Ohishi}}, \bibinfo {author} {\bibfnamefont
  {Y.}~\bibnamefont {Kousaka}}, \bibinfo {author} {\bibfnamefont
  {J.}~\bibnamefont {Akimitsu}}, \bibinfo {author} {\bibfnamefont
  {K.}~\bibnamefont {Inoue}},\ and\ \bibinfo {author} {\bibfnamefont
  {S.}~\bibnamefont {Ohara}},\ }\bibfield  {title} {\bibinfo {title} {Neutron
  diffraction study of antiferromagnetic $\rm{ErNi_3Ga_9}$ in magnetic
  fields},\ }\href {https://doi.org/10.1016/j.physb.2017.09.057} {\bibfield
  {journal} {\bibinfo  {journal} {Physica B}\ }\textbf {\bibinfo {volume}
  {536}},\ \bibinfo {pages} {392} (\bibinfo {year} {2018})}\BibitemShut
  {NoStop}%
\end{thebibliography}%

\end{document}